# Relationship between inflation, unemployment and labor force change rate in France: Cointegration test

I.O. Kitov, IDG RAS, ikitov@mail.ru

O.I. Kitov, the University of Warwick, O.Kitov@warwick.ac.uk

S.A Dolinskaya, IDG RAS

## Abstract

A linear and lagged relationship between inflation, unemployment and labor force change rate, $\pi(t)=A_0UE(t-t_0)+A_1dLF(t-t_1)/LF(t-t_1)+A_2$ (where $A_0$, $A_1$, and $A_2$ are empirical country-specific coefficients), was found for developed economies. The relationship obtained for France is characterized by $A_0$=-1, $A_1$=4, $A_2$=0.095, $t_0$=4 years, and $t_1$=4 years. For GDP deflator, it provides a root mean square forecasting error (RMFSE) of 1.0% at a four-year horizon for the period between 1971 and 2004.

The relationship is tested for cointegration. All three variables involved in the relationship are proved to be integrated of order one. Two methods of cointegration testing are used. First is the Engle-Granger approach based on the unit root test in the residuals of linear regression, which also includes a number of specification tests. Second method is the Johansen cointegration rank test based on a VAR representation, which is also proved to be an adequate one via a set of appropriate tests. Both approaches demonstrate that the variables are cointegrated and the long-run equilibrium relation revealed in previous study holds together with statistical estimates of goodness-of-fit and RMSFE.

Relationships between inflation and labor force and between unemployment and labor force are tested separately in appropriate time intervals, where the Banque de France monetary policy introduced in 1995 does not disturb the long-term links. All the individual relationships are cointegrated in corresponding intervals.

The VAR and vector error correction (VEC) models are estimated and provide just a marginal improvement in RMSFE at the four-year horizon both for GDP deflator (down to 0.9%) and CPI (~1.1%) on the results obtained in the regression study. The VECM approach also allows re-estimation of the coefficients in the individual and generalized relationship between the variables both for cointegration rank 1 and 2.

Comparison of the standard cointegration approach to the integral approach to the estimation of the coefficients in the individual and generalized relationships between the studied variables demonstrates the superiority of the latter. The cumulative inflation curve or inflation index, which is the actually measured evolution of price level, is much better predicted in the framework of the integral approach, which is a powerful tool for revealing true relationships between non-stationary variables and can be potentially used for rejection of spurious regression. The cumulative curves allow avoiding obvious drawbacks of the VECM representation and cointegration tests – increasing signal to noise ratio after differentiation and severe dependence on statistical properties of error terms.

The confirmed validity of the linear lagged relationship between inflation, unemployment and labor force change indicates that since 1995 the Banque de France has been wrongly applying the policy fixing the monetary growth to the reference value around 4.5%. As a result of the policy, during the last ten years unemployment in France was twice as large as the one dictated by its long-term equilibrium link to labor force change. This increased unemployment compensates the forced price stability.

**Introduction**

A linear lagged relationship between inflation, unemployment and labor force change has been obtained for several developed countries (Kitov, 2006ab, 2007). For France, this relationship is characterized by a high predictive power and explains more than 90% of variability in GDP deflator. It covers the period of continuous measurements between 1971 and 2006 providing a root mean square forecasting error (RMSFE) close to 1% at a four-year horizon. (The first paper for France (Kitov, 2007) was finished in 2005 and thus the period ends in 2004.) Shorter forecast horizons are characterized by the same accuracy, i.e. the four years between a change in labor force and the reaction of corresponding inflation do not add any information to the current value for inflation. Effectively, no processes or phenomena during this four-year period can change the future inflation.

In the USA, Japan, and Austria the general relationship between these three variables can be split into two separate dependencies of inflation and unemployment on labor force change with their own coefficients and time lags. Such a split is possible because of the absence of any artificial constraints on both inflation and unemployment in the USA and Japan. Fortunately for Austria, monetary constraints of the European System of Central Banks (ESCB) almost completely correspond to the long-run equilibrium evolution of inflation and unemployment as linear function of labor force change. For France, however, the same monetary constraints have induced a very strong deviation from the natural evolution of unemployment as defined by the long-run dependence on labor force change (Kitov, 2007). These constraints are formulated in the ECB (2004) documents as related to price stability requirement. Banque de France (2004) explicitly defines corresponding target value:

> **The reference value for monetary growth must be consistent** with—and serve—**the achievement of price stability.** Furthermore, the reference value for monetary growth must **take into account real GDP growth and changes in the velocity of circulation of money.** The derivation of the reference value is based on the contributions to monetary growth resulting from the achievement of the ultimate objective of price stability (year-on-year increases of below 2%), and from the assumptions made for potential GDP growth (2-2.5% per annum) and the velocity of circulation (a trend decline of about 0.5-1% each year).

Taking account of these two factors, **the Governing Council decided to set the first reference value at 4.5%.**

Effectively, this reference value of 4.5% creates an artificial barrier in the French economy not allowing natural interaction of macroeconomic parameters. In particular, the labor force started to grow in the middle 1990s. According to the long-run equilibrium dependence on labor force change rate, inflation had to increase to the rate between 5% and 10% per year, and unemployment had to decrease to the level around 4%. The monetary barrier did not allow this scenario, however, and the potential inflation growth has been channeled through an elevated unemployment (Kitov, 2007). Despite the strong deviation in each of the individual dependencies, the generalized relationship between inflation, unemployment and labor force holds before and after 1995. This generalized equation accurately predicts inflation at a four-year horizon as regression analysis demonstrates.

All three variables involved in the relationship are non-stationary, implying a possibility for the regression results to be spurious despite the existence of a theoretical foundation (Kitov, 2006b, 2007). Therefore, econometric tests of the relationship for cointegration are necessary. If the studied relationship is a cointegrating one, the results of the previous regression analyses hold from the econometric point of view. One has to bear in mind, however, that modeling inflation as a function of labor force change rate is associated with the risk to obtain biased results when econometric methods are applied without detailed consideration of error sources (I.Kitov, O.Kitov & Dolinskaya, 2007).

The remainder of the paper is organized in four sections. Section 1 briefly introduces the model and presents some results obtained for France in the previous study. Section 2 is devoted to the estimation of the order of integration in measured inflation, unemployment and labor force change rate. Unit root tests are carried out for original series and their first differences. GDP deflator and CPI represent inflation in the study.

In Section 3, the existence of a cointegrating relation between three variables is tested. The presence of a unit root in the difference between the measured and predicted inflation implies an absence of cointegration between the variables and a strong bias in the results of the previous regression analysis. The residuals obtained from regressions of the measured inflation on the predicted ones are also tested for the unit root presence.

This approach is in line with that proposed by Engle and Granger (1987). The maximum likelihood estimation procedure developed by Johansen (1988) is used to test for the number of cointegrating relations in a vector-autoregressive (VAR) representation. The existence of a cointegrating relationship is studied as a function of the predictor smoothness – from the original predictor to its three-year moving average. The VAR and vector-error-correction (VEC) models are estimated for forecasting purposes.

Section 4 discusses relative advantages and drawbacks of two approaches to finding valid macroeconomic relationships: cointegration and integration. The first is based on the representation of actual non-stationary observations in the form of lagged differences - VAR or VECM. The underlying assumption of the first approach is the presence of independent stochastic trends in the data. The second approach assumes a true link between involved variables and uses integrative methods adopted from physics.

Section 5 discusses principal results and their potential importance for economic theory and practical application in France.

**1. The model and data**

France is characterized by an outstanding productivity and has the largest GDP per working hour among large developed economies, as presented by the Conference Board and Groningen Growth and Development Center (2005). At the same time, real economic performance in France has been far from stellar during the last twenty-five years with a mean annual real GDP growth of 2%. Therefore, France is an example of an economy different in many aspects from those in the USA, Japan, and Austria. This is especially important for the concepts we examine.

Original data for France have been obtained from the OECD web-site (2006), which provides time series of various lengths for the variables involved in the study: GDP deflator (*GDPD*), CPI (*CPI*), labor force level (*LF*), and unemployment rate (*UE*). Similar series are also available from different sources such as INCEE (http://www.insee.fr) and Eurostat (http://epp.eurostat.ec.europa.eu). In some cases, a large discrepancy between supposedly identical series is observed (Kitov, 2007).

There are two different measures of inflation in France used in this study: GDP deflator and CPI. In general, they are similar with only relatively small discrepancies during some short intervals but give fairly different statistical results. GDP deflator is probably the best measure of inflation to model. It adequately reflects inherent links between price increase and labor force growth, as found for the USA, Japan, France, and Austria (Kitov, 2006ab, 2007). Both the *CPI* and *GDPD* demonstrate a high inflation rate between 1975 and 1985 and a gradual decrease to the current level close to 2%. The CPI time series is used to extend the inflation model to the late 1960s, where there are no GDP deflator measurements available.

The start of the current period of labor force growth almost coincides with the establishment of a new entity of the French national bank, Banque de France, as an independent monetary authority having a fixed target value of inflation rate. In 1993, the ESCB dramatically changed its approach to inflation managing – the main target is now to reach price stability at a level near 2% of annual growth (ECB, 2004).

For France, we use similar to the previous study procedure to fit annual and cumulative inflation and unemployment readings by linear functions of labor force change rate. The cumulative curve is most sensitive to coefficients in the relationship between measured and predicted variables. Even a small systematic error in predicted amplitude cumulates to a high value when aggregated over thirty-five years. The procedure results in the following relationship between unemployment, *UE(t)*, and labor force change rate, *dLF(t)/LF(t)* (Kitov, 2007):

$$UE(t) = 0.165 - 13dLF(t)/LF(t) \qquad (1)$$

Linear coefficient in (1) amplifies labor force change and correspondingly any measurement error in the labor force by a factor of 13. This coefficient is also a negative one, i.e. any increase in labor force is converted into a synchronized (no time lag between the labor force and the unemployment change) drop in unemployment rate in France. On the other hand, in the absence of any growth in the labor force the unemployment rate reaches a 16.5% level.

A standard linear regression analysis of relationship (1) is carried out for the period between 1970 and 1995. The measured unemployment time series is characterized by stdev=0.032. The regression gives $R^2$=0.48 with root mean square error (RMSE) of 0.023. The annual time series is poorly predicted. Smoothing of the labor force and unemployment series by moving average results in a significant improvement in the goodness-of-fit and standard deviation (Kitov, 2007).

Inflation is also represented by a linear function of labor force change but with a large lag. For the GDP deflator, the following relationship is obtained (Kitov, 2007):

$$GDPD(t) = 17dLF(t-4)/LF(t-4) - 0.065 \qquad (2)$$

where *GDPD(t)* is the inflation at time *t*, *LF(t-4)* is the labor force four years earlier. Thus, there is a four-year lag in France between the labor force change and corresponding reaction of the inflation. The true lag is around 4.5 years because of the difference between timing of inflation and labor force measurements. The former corresponds to the last day of year as the price change accumulated during the year. The latter actually represents the averaged value of monthly readings and should better fit the measurement in July than the one in December. One can chose between four- and five-year lag. Our choice is the four-year delay.

The value of the linear coefficient (17) indicates that the inflation is also very sensitive to the labor forced change. The intercept -0.063 means that a positive labor force change rate has to be retained in order to avoid deflation. The threshold for a deflationary period is a change rate of the labor force of 0.0037(=0.063/17) per year.

Since the discrepancy between the observed and measured inflation starts in 2000, a linear regression analysis is carried out for the period between 1971 and 1999. The GDP deflator is a dependent variable and a predictor is obtained according to relationship (2). Standard deviation of the actual time series for the studied period is 0.042. The regression of the annual readings is characterized by $R^2$=0.47 and stdev=0.031.

The discrepancy between the observed and predicted time series started in 1996 for the unemployment and 2000 for the inflation. It is explained by the new monetary policy first applied by the Banque de France in the middle 1990s. The policy of a constrained money supply, if applied, could obviously disturb relationships (1) and (2).

Our explanation of the inflation and unemployment reaction on the change in the monetary policy in France is as follows. Money supply in excess of that related to real

GDP growth is completely controlled by the demand of growing labor force, because the excess is always accommodated in a developed economy through employment growth, which causes inflation. The latter serves as a mechanism which effectively returns personal income distribution, normalized to total population and nominal GDP per capita growth, in the economy to its original shape (Kitov, 2006ab). The relative amount of money that the economy needs to accommodate a given relative labor force increase through employment is constant through time in the corresponding country but varies among developed countries. This amount has to be supplied to the economy, however. Central banks are responsible for this process. In the USA and Japan, central banks provide adequate procedures for money supply and individual dependence on labor force change does not vary with time both for inflation and unemployment. The ESCB limits money supply to achieve price stability. In Austria, it does not affect the individual linear relationships because the actual money supply almost equals the amount required by the observed labor force growth. For France, the labor force growth is so fast that it demands a much larger money input for the creation of an appropriate number of new jobs. The 2% artificial constraint on inflation (and thus money supply) disturbs relationships (1) and (2). The labor force growth induces only an increase in employment, which accommodates the given 2% inflation instead of the 9% predicted inflation. Those people who enter the labor force in France in excess of that allowed by the target inflation have no choice except to join "the army of unemployed". Hence, when inflation is fixed, the difference between observed and predicted change in the inflation must be completely compensated by an equivalent change in unemployment in excess of the predicted one. The generalized relationship (3) mathematically describes this assumption.

For France, this generalized relationship is obtained as a sum of (1) and (2), which gives the following equation:

$$GDPD(t) = 4dLF(t-4)/LF(t-4) - UE(t-4) + 0.095 \quad (1973<t<2004) \quad (3)$$

where the intercept 0.095 is slightly different from that obtained as a straight sum of corresponding free terms: 0.165-0.063=0.102. The difference is dictated by the fit of the cumulative curves. It is important to emphasize that relationship (3) is valid for the entire interval where the OECD GDP deflator readings are available, except may be the first two years of the period due to the four-year delay of inflation reaction on labor force change, as explained later on in the paper.

A linear regression of the observed inflation against that predicted according to (3) is characterized by $R^2$=0.88 and stdev=0.014, which is remarkable for annual curves. These values are the best we have obtained for France so far in this paper. They explain the inflation to the extent beyond which measurement uncertainty should play a key role. Practically, there is no room for any further improvements in $R^2$ given the current time series. The regression results also undoubtedly prove the success of the generalized approach.

As a result, we have obtained a very accurate description of inflation in France during the last 35 years. In contrast to Austria, a prediction of inflation for the next four years can be computed using only past readings of the labor force. No population projections are necessary for the inflation forecast at the four-year horizon. At longer horizons, one can use labor force forecasts. Accuracy of such long-term unemployment and inflation forecasts is proportional to the accuracy of the labor force predictions. Monetary policy of the ECB is not an important factor for the forecast despite its influence on the partition of the labor force growth between inflation and unemployment. The sum of these two variables is always a linear function of the labor force change, however. Therefore, it is for the ECB and Banque de France to decide on the future partition of the labor force growth into unemployment and inflation. There is no opportunity to compensate the past high unemployment by freeing monetary supply, however. To achieve the predicted 4% unemployment rate a further fast growth in the labor force is necessary. Otherwise, the unemployment will be maintained at its current level.

It is clear from the behavior of the measured time series, as displayed in Figure 1a, that all three variables (inflation is represented by GDP deflator and CPI) are potentially characterized by the presence of unit roots. In such a situation a spurious regression is probable as modern econometric research shows (Granger & Newbold, 1967). Therefore, some specific tests have to be carried out in order to validate the results obtained by Kitov (2007). In particular, one has to prove that the measured time series are integrated of order one, I(1), and are cointegrated, i.e. the residuals of their regression create a stationary time series.

A similar analysis has been carried out for the relationship between inflation and labor force change rate in the USA (I.Kitov, O.Kitov & Dolinskaya, 2007). This study has shown the existence of a cointegrating relation. There is a significant complication of the analysis in the case of France, however. The USA analysis was essentially a bivariate one. For France, a trivariate analysis is mandatory because of the deviations from the individual long-run relationships starting in 1996 for unemployment and 2000 for inflation. The deviations forced the usage of generalized relationship (3) linking all three variables in one equation. Therefore, additional efforts are necessary for determination of cointegration rank in the framework of VAR methodology. The trivariate analysis does not dismiss the possibility to test the individual relationships for cointegration during the periods where they hold. The individual analysis might be also of practical interest if the French monetary authorities abolish the current unsound policy in future.

**2. The unit root tests**

We start with unit root tests for the measured series for determination of the order of integration. If the time series are I(1), they have to be characterized by the presence of a unit root and their first difference by the absence of unit roots (Hendry & Juselius, 2001). The augmented Dickey-Fuller (ADF) and the Dickey-Fuller general least-squares (DF-GLS) tests from a standard econometric package Stata9 are used in this study.

There are several series modeled by Kitov (2007) between 1971 and 2004 which have to be tested for the unit root presence. Except for the *GDPD*, the original time series span longer time intervals. CPI estimates are available after 1956. Labor force change rate, *dLF/LF,* and unemployment span the period after 1957. The latter time series can be divided into a number of segments with various units of measurements as described by OECD (2005). As a rule, new definitions of unemployment included more people. This makes any statistical estimates carried out for the unemployment time series less reliable. It is worth noting that unemployment is a complimentary part of labor force and introduces a proportional disturbance in the latter. The net effect of the revisions in unemployment definition on the labor force readings is relatively low but potentially results in a bias.

Table 1 reports some descriptive statistics for the original series and their first differences. The mean values of the differences are close to zero. Corresponding curves shown in Figure 1b obviously demonstrate fluctuations around a zero mean. Since the variables are apparently non-stationary, the empirical frequency distributions are not normal. Considering the results in Table 1, we accept the assumption that the first differences (*dGDPD*, *dCPI*, *dUE*, and *d(dLF/LF)*) have zero means, and hence, corresponding original series contain no linear trends. This observation is important for the specification of unit root tests.

Table 2 lists results of the ADF and DF-GLS tests for a unit root in the four measured time series. Despite the previous analysis of relationship (3), which was limited to the period after 1973, it is instructive to use the entire time series for the unit root tests. This increases the number of readings by a factor of 1.5, and provides more robust statistical estimates. For the unemployment, one might expect deterioration in statistical inferences due to unreliable readings before 1971. Inclusion of any lag results in test values well above the corresponding 1% critical value given in brackets. The worst two cases are in the *GDPD* due to the shortness of the series and the unemployment. The latter effect is expected due to the construction of the series. The *CPI* series practically repeats that of GDP deflator between 1971 and 2004. Therefore one can presume similar results for the *GDPD* as for the *CPI* if the former would be extended to 1957.

The principal conclusion from Table 2 is that no one of the four series is a stationary one because, the null hypothesis of the unit root presence can not be rejected. The closest case to the rejection of the null hypothesis is for the series *dLF/LF,* which practically represents the first difference of the labor force level. The ADF test with the maximum lag 0 even rejects the null hypothesis.

The variables in Table 2 are non-stationary. But it does not mean that the series are I(1) and additional efforts are necessary to prove the assumption. For an I(1) process, the first difference has to be an I(0). Therefore, the same tests are repeated on the first differences of the studied series. Results are presented in Table 3.

A standard unit root test usually contains many specifications related to such statistical features as heteroskedasticity, autocorrelation, non-normality and others, which potentially present in a time series. Despite the zero means in the first differences, the tests listed in Table 3 use two different assumptions on the presence of a constant and a linear trend. We test for a single unit root in an AR(p) representation, where p=0,…,3 for the ADF test and p=1,…,4 for the DF-GLS test. The choice of the highest lag in the AR representation is a difficult problem bearing in mind that we only have from 33 to 47 readings in the series. Auto-correlograms show the absence of any significant correlation at lags beyond four years. The largest autocorrelation is observed at lag 1 and sometimes at lag 3. For the *dLF/LF*, there is a significant (>0.2) autocorrelation at lags around 10. We consider them as artificial and related to side effects. Therefore we limit the AR model to the largest lag 4, with test results being unreliable at this and larger lags.

When a constant term is allowed in the series, the ADF tests reject the null hypothesis of a unit root in the labor force change rate and CPI for any lag, except for lag 4 for the *dLF/LF* series. The DF-GLS tests provide similar results but for the largest lag 3 and 2 for the *CPI* and *dLF/LF*, respectively. When the trend term is included, the tests produce very similar results, except that for the ADF for the *CPI*, where the null hypothesis is not rejected for the highest lag 4. Therefore, we reject the presence of unit roots in the first difference of the labor force change rate and CPI series. Hence, the original series are I(1).

When a constant term is used in the first difference of the GDP deflator series, the ADF test rejects the null only for the maximum lag 0 and 2; the DF-GLS test rejects the null for the highest lag 0 and 1. With a linear trend included, both the ADF and the DF-GLS tests accept the null hypothesis at all lags. The *GDPD* series practically coincides with the *CPI* one, however. Thus, one can expect the same result in the tests when the GDP series is extended to the size of the CPI series, and we assume that the GDP deflator series is integrated of order one as well.

The DF-GLS test with a constant tern rejects the presence of a unit root in the first difference of the unemployment series at any lag. The ADF test rejects the null at lags zero and one. When a linear trend is included, both tests accept the null hypothesis. The trend should not be included as a clear misspecification of the tests. As for the other three variables, we treat the *UE* as an I(1) time series and are ready for cointegration tests.

## 3. The cointegration tests

At first, we estimate an unrestricted VAR(2) model with a constant term for the three studied variables: *GDPD*, *UE*, *dLF/LF*. For obvious reasons, the latter two are considered as weakly exogenous variables and are shifted by four years ahead in order to synchronize them with the GDP deflator. There are only 32 readings between 1973 and 2004, so the model might be unreliable.

We have failed to extend relationship (2) to the period before 1969. Accordingly, we have to move the start point of the modeled period from 1971 to 1973 because of the four-year lag of the inflation behind the labor force change and unemployment.

The highest lag recommended by the lag-order selection statistics (LR, FPE, AIC, HQIC, SBIC) is 1 for the pre-estimation and 2 for the post-estimation of the VAR model. The Lagrange multiplier (LM) test does not reject the null hypothesis of no autocorrelation for lags between 1 and 4. The residuals of the *GDPD* in the model are characterized by skewness=0.35 and kurtosis=2.64. Therefore, the Jarque-Bera test does not reject the null hypothesis of normality. All eigenvalues of the model lie inside unit circle (0.59, 0.59) - the VAR model satisfies a stability condition. The roots are not close to unity indicating that the residual series of the VAR is stationary. This evidences in favor of cointegration. The observed non-stationarity in the series is driven by the two exogenous variables. Hence, the VAR model provides an adequate description of the studied processes.

When all three variables are considered as endogenous in a VAR model, the highest lags recommended by various lag-order selection statistics are as follows: 4 (LR), 2 (FPE), 4 (AIC), 2 (HQIC), and 1 (SBIC). A reasonable lag order would be 3. The LM test also assumes the acceptance of the null hypothesis of no autocorrelation in the VAR(3) model with endogenous variables. The Jarque-Bera test for normality in the VAR residuals rejects neither of the nulls for the three variables separately and jointly. The largest eigenvalue in the VAR(3) model is 0.96, with the closest two roots 0.83 being equal. Because the largest root is very close to unity the series are probably non-stationary and one can expect the existence of a single cointegrating relation. In practice the apparent misspecification of the *dLF/LF* and *UE* as endogenous variables relative to the *GDPD* results in the necessity to reformulate the VAR model into a vector error correction model due to the potential presence of a unit root.

Before building a VECM for the variables we analyze the residuals of relationship (3). The assumption that inflation, unemployment and labor force change rate in France are three endogenous, non-stationary and cointegrated time series is equivalent to the assumption that the difference, $\varepsilon(t)$, between the measured, $\pi_m(t)$, and predicted, $\pi_p(t)$, inflation: $\varepsilon(t)=\pi_m(t)-\pi_p(t)$, is a stationary series. (In relationship (3), the left-hand side term is the measured inflation, and the right-hand side term is the predicted inflation.) A natural next step in this case is to test the difference for the unit root presence. If $\varepsilon(t)$ is a non-stationary variable having a unit root the null of the existence of a cointegrating relation can be rejected. Such a test is associated with the Engle-Granger's approach, which, however, requires $\pi_m(t)$ to be regressed on $\pi_p(t)$ at the first step.

The hypothesis for a unit root is tested by the same procedures and with the same specifications as used in Section 2. If the null hypothesis of a unit root is rejected, the hypothesis of no-cointegration is also rejected. In this case, the equilibrium relationship (3) between the measured and predicted inflation is valid and a vector error-correction model can be estimated using results of the first stage. We test several differences. The measured series is represented by the *GDPD*, and the predicted series are presented by the original one and those smoothed by a moving average: $diff1=\pi_m(t)-\pi_p(t)$, $diff2=\pi_m(t)-MA(2)$, $diff3=\pi_m(t)-MA(3)$, $diff4=\pi_m(t)-MA(4)$. The original predicted time series is characterized by very high fluctuations reflecting the high sensitivity of the inflation and unemployment to the labor force change rate. The original series and *diff1* are displayed in Figure 2a and 2b, respectively. For the *diff1,* probability for the presence of a unit root is high. At the same time, the usage of the moving averages as predictors is an absolutely valid operation because these averages include only those values of the labor force and unemployment, which are in the past relative to the modeled inflation readings. For France, the moving averages provide better forecasts and an adequate description of inflation (Kitov, 2007).

Figures 3 through 5 provide the measured, predicted from the labor force change rate, and related residual time series for the GDP deflator, CPI, and unemployment. Corresponding unit root tests are carried out but are not discussed because of their principal similarity to the trivariate approach.

Table 4 presents some results of the unit root tests. As expected, the ADF and DF-GLS results indicate the presence of a unit root in the *diff1* residual series, except lags 2 and 3 for the ADF tests. For the series *diff2* through *diff4*, the null hypothesis of a unit root is consistently rejected by the DF-GLS tests and only the ADF test without lagged differences accepts the null. One can conclude that the residuals obtained from relationship (3) build a stationary time series and the observed and predicted inflation are cointegrated.

The next step is to use the Engle-Granger approach and to study statistical properties of the residuals obtained from a linear regression of the *GDPD* on *UE* and *dLF/LF* series. We consider the latter two variables as exogenous in the regression. This approach is similar to first of the above VAR models but does not use any past values of the inflation to describe its current value. Table 5a summarizes results of some specification tests for the regression residuals, constants with their standard deviations and t-test results, $R^2$, RMSFE. The dependent variable is always the *GDPD* and predictors vary from the trivariate model (*UE+dLF/LF*) to *MA(3)* of the predicted inflation.

Despite high $R^2$ values, results of the specification tests indicate that the residuals of the regressions are characterized by heteroskedasticity, autocorrelation and deviation from normality. In addition, the residual time series definitely contains a unit root as the ADF and DF-GLS tests show. The best RMSFE in Table 5a is only 0.012 compared to 0.0095 provided by the VAR. The best RMSFE obtained by Kitov (2007) is 0.01. So, the VAR model with two exogenous variables statistically better describes the link between the variables.

Similar specification tests are fulfilled for the GDP deflator between 1971 and 1999, for the CPI between 1967 and 1999, and for the unemployment between 1971 and 1995; all variables are regressed on the labor force change rate. Tables 5b through 5d demonstrate that the regressions provide residuals distributed close to normality, without heteroskedasticity and omitted variables. Autocorrelation is also low in the residual series.

Johansen (1988) approach is based on the maximum likelihood estimation procedure and tests for the number of cointegrating relations in the vector-autoregressive representation. This allows simultaneous tests for the existence of cointegrating relations and determining their number (rank). For three variables, one or two cointegrating relations are possible. In the case of France, the estimation of the number of cointegrating relations is complicated by the dramatic change in monetary policy around 1995. This decision divides the period between 1973 and 2004 into two unequal segments. The first segment is characterized by the presence of two individual relationships (1) and (2), which hold independently. After 1995, only relationship (3) holds and the individual dependencies are strongly disturbed. Therefore, the Johansen test for cointegration rank will probably give split results depending on specification.

Table 6a lists trace statistics, eigenvalues, LL obtained from the cointegration rank tests in the trivariate model. There are three different specifications of deterministic terms tested - a constant in the time series (constant), a constant in cointegrating relation (rconstant), and no deterministic term (none). The maximum lag order has been varied from 1 to 4. As expected, the tests demonstrate mixed results. Dominating cointegration rank is 1. There are four cases of rank 0, for the maximum lag order 3 for all three specifications, and one case indicates cointegration rank 2. Our best assumption is the existence of a constant term in the time series. Hence, we accept the existence of a single cointegrating relation between the GDP deflator, unemployment and labor force change rate in France between 1973 and 2004.

The Johansen tests for the observed and predicted inflation give cointegration rank 1 for the maximum lag order 3 and 4 in both studied specifications. Lag 1 produces rank 1 for the constant term in the series and rank 0 for the constant term in cointegrating relation. In the series predicted according to relationship (3), relative inputs of the labor force and unemployment is fixed. This drawback (because of less degrees of freedom) compared to the trivariate model results in a larger RMSFE. But even in this disadvantageous situation cointegrating relation does exist.

Tables 6b through 6d report results of cointegration rank tests for the GDP deflator, CPI, and unemployment, where the variables are represented by individual linear lagged functions of the labor force change rate in corresponding periods. These results are illustrative for statistical tests carried out at small time series. Some cointegration tests for the GDP deflator between 1971 and 1999 result in the absence of any cointegration between the predicted and measured time series, although an extension of the series by several readings, as presented by the CPI, results in the acceptance of the cointegration rank 1 for any maximum lag between 1 and 4 and the constant deterministic term in the original series.

Another important observation is the change in the estimates of cointegration rank with increasing smoothness of predictor. When a moving average replaces original predicted time series, the rank consistently drops to 0, i.e. indicates the absence of cointegration. This effect is related to the "worsening" of auto-correlative properties of the moving averages. As discussed in (I.Kitov, O.Kitov & Dolinskaya, 2007), the increasing accuracy of prediction due to suppression of random errors is accompanied by the increasing influence of systematic errors destroying fundamental assumptions of econometric approach. In other words, the better one can predict some macroeconomic variable, the poorer statistical inference s/he obtains. For a scientist, the situation is easily resolved by the improvement in measurement accuracy in further experiments. For an econometrician, solution is not so easy due to the overpressure of huge theoretical legacy. To some extent, professionals in economics and econometrics are not interested in obtaining an appropriate level of measurement uncertainty of macroeconomic variables. There is a probability that conventional econometric approach to macroeconomics might be destroyed with increasing measurement accuracy. This consideration does not affect very well measured micro-economic and financial time series. As in physics, fundamental conservation laws valid for a closed system as a whole do not deny a possibility of large fluctuations in small sub-volumes.

Because of the four-year lag behind the labor force change, the two-year recommended lag in the VAR model and the existence of cointegrating relation(s) or long-run equilibrium relationship(s), a test on causality direction is a redundant one. It is obvious that the labor force change rate and unemployment series are weakly exogenous for the inflation.

We have carried out several formal cointegration tests for the relationships between inflation, unemployment and labor force change rate and obtained an overall confidence in the existence of a true linear and lagged link between the variables. Is it possible, however, to extend our study in this specific case and to increase our confidence?

**4. Cointegration analysis or cumulative curves?**

The existence of an equilibrium long-run relationship between several stationary or non-stationary variables also implies that this link holds for derivatives and integrals of the variables. Preservation of a true link between derivatives or first differences is extensively used in the concept of cointegration. The explicit idea behind the cointegration approach consists in removing any type trends, stochastic or deterministic, from non-stationary time series. When the trends are stochastic and independent for the variables involved in the link, differentiation effectively suppresses the influence of exogenous forces acting differently at the variables and retains the true link. There is some doubt, however, that the differentiation is a good method to reveal the link. First, the exogenous forces causing the stochastic trends partly retain their influence in the first differences biasing statistical estimates. Second, the true link between the original non-stationary variables defines a common deterministic trend, which is removed by differentiation.

Due to a lower relative uncertainty usually associated with measurements of integral values, the existence of a strict link between integrals can be used for a more accurate estimate of coefficients in cointegrating relationship. Such an integral-based or cumulative approach is especially important for those variables, which are actually measured as levels in economics. The purpose of this Section is to demonstrate that the integral technique is superior to the cointegration analysis in revealing true links between non-stationary time series and estimation of their coefficients.

For the sake of simplicity, we call the curves consisting of annual readings the "dynamic" curves and those obtained by progressive summation of the annual readings the "cumulative" ones. The latter curves can be also called "integral" curves. This nomenclature differs from that accepted in econometrics, where the term "dynamic" is usually associated with the first differences.

So, we have an assumption of the existence of a strict dynamic relationship between some variables. Unfortunately, the assumption can not be validated by standard methods associated with the improvement in relevant measurement accuracy. In such a situation only indirect procedures highlighting specific aspects of the relationship and its error term are possible. Cointegration tests look through a magnifying glass at the error term and allow a judgment on the presence or absence of the dynamic relationship itself. A fundamental assumption in any cointegration test is that the error term has to be i.i.d. (independent and identically distributed) with zero mean and finite variance that is not always the case in macroeconomics. As a result, cointegration tests are an effective tool for rejection of both spurious regressions and valid relationships (Chiarella & Gao, 2002).

The approach using cumulative curves takes an advantage of the increasing relative accuracy of integral values, when the latter are the actually measured values such as price, labor force and unemployment levels. If a true link between the variables does exist, the error term in the integral relationship has to be as statistically good as the error term associated with the dynamic representation. Cumulative values, i.e. the net change between initial and current measured levels, are estimated with increasing relative accuracy, however, and the relative error term evolves in inverse proportion to the net change in corresponding level. (For example, one could measure an average speed of a car more accurately using a ratio of total distance and time than integrating instantaneous speed measurements.) Therefore, the integral approach provides a powerful tool for discrimination between valid and spurious relationships. Moreover, the coefficients in the relationship obtained by minimizing the difference between cumulative curves are superior to those obtained by linear regression and in the VECM representation. In the best case, the latter coefficients provide an error term defining a random walk for cumulative curves, which is characterized by growing variance with increasing length of time series.

Theoretically, the cointegrating relationship between *GDPD*, *UE* and *dLF/LF*, i.e. between change rates of actually measured (level) variables, implies that the error term is represented by white noise. In an ideal case for the OLS estimation, the noise has a normal distribution, $N(0,\sigma^2)$. This requirement is a weak point of cointegration analysis. It is aimed to find such coefficients in a cointegrating relation, which provide the desired error distribution. This is not enough for an adequate description of the relationship between the integrals of the variables, however. If the error term is a random innovation from the $N(0,\sigma^2)$ distribution, then the cumulative value of the error term will not guarantee the convergence of the cumulative curves

$$\Sigma GDPD(t_i) = \Sigma(A_1 dLF(t_i-t_1)/LF(t_i-t_1) - A_2 UE(t_i-t_2) + A_3) + \Sigma\varepsilon_i,$$

since the standard deviation of the random walk process, $\Sigma\varepsilon_i$, increases proportionally to the square root of the series length, $T - N(0, \sqrt{T},\sigma^2)$. This means that the increasing discrepancy between cumulative curves is very probable if the cointegrating relation is obtained in the VECM or VAR approach. To provide an adequate description of the cumulative curves one has to keep the integral value of the error term fluctuating around zero mean, i.e. to guarantee a quasi-white noise distribution for the integral error term. In practice, measuring procedures for such economic parameters as labor force level and unemployment contain so many artificial procedures and revisions, which change past values, that one can not expect measurement noise even close to white one.

A linear regression analysis of the link between the *GDPD*, *UE* and *dLF/LF* has been carried out by Kitov (2007). In the VECM representation, coefficients of cointegrating relations (with the imposed Johansen normalization restriction) and related standard errors have been obtained for various lags and ranks. Results are listed in Table 7a. We have also estimated several VEC models for the unemployment and GDP deflator as functions of *dLF/LF*. Corresponding coefficients are presented in Table 7b. The cointegrating relations in Tables 7a and 7b are usually different from those obtained from simple linear regressions due to the inclusion of additional parameters describing the true link. As a rule, the increased number of parameters has to provide a more accurate approximation.

For cointegration rank 1 in Table 7a, i.e. in the case of the simultaneous estimation of coefficients in a single cointegration relation, the slope associated with *UE* decreases from -0.457 for the maximum lag 1 to values near -1.0 (-1.05 for lag 3 and -0.957 for lag 4). In relationship (3) and (6) we have fixed the *UE* slope to -1.0 by definition. When the *UE* slope is close to -1.0, the *dLF/LF* slope fluctuates between 3.12 and 5.11. The variation of the slopes is compensated by a changing constant term.

For cointegration rank 2, there is an obvious trade-off between the sum of the slopes associated with *dLF/LF* and *UE* and the intercept term in Table 7a. The sum varies from 2.43 for the maximum lag order 4 to 5.25 for the maximum lag 3. At the same time, the wide range of the slopes' variation (from 18.90 to 64.91 for *dLF/LF* and from -14.92 to -62.48 for *UE*) demonstrates the inconsistency of the assumption of two cointegrating relations between the three variables over the whole period between 1973 and 2004. According to Kitov (2007), the last eight years are characterized by individual relationships different from (1) and (2). Hence, the coefficients obtained in the VECMs might be strongly biased.

For illustration of the discrepancy between cumulative curves when coefficients are obtained in the framework of the VECM approach we have chosen several cases from Tables 7a and 7b. For the GDP deflator as a linear lagged function of *dLF/LF*, a VECM with one cointegrating relation and the maximum lag 2 gives the following relationship:

$$GDPD(t) = 17.45[0.85]dLF(t-4)/LF(t-4) - 0.063, \qquad (4)$$

which is very close but different from (2). This difference may seem marginal for the dynamic (i.e. annual rates) curves *GDPD* and *dLF/LF*. Relationship (2), however, was obtained using cumulative curves, which are very sensitive to the free term - even an error of 0.001 would give 3.5% deviation in 35 years. Figure 6 demonstrates the failure of relationship (4) to predict the long-term evolution of the GDP deflator index. Therefore, coefficients in (4), minimizing the distances between the measured and predicted annual curves, do not provide the lowermost average distance between corresponding cumulative curves.

For the unemployment, a VECM with the maximum lag 2 gives the following relationship:

$$UE(t) = -11.97[1.20] \, dLF(t)/LF(t) + 0.157, \qquad (5)$$

which is also close to relationship (1). Figure 7 depicts the deviation between the cumulative curves defined by (1) and (5). Visually, the deviation is a minor one, but quantitatively is measurable.

Using (4) and (5) one can rewrite the generalized relationship (3) in the following form:

$$GDPD(t) = 5.48 \, dLF(t-4)/LF(t-4) - UE(t-4) + 0.094. \qquad (6)$$

This is a cointegrating relation between the three variables as obtained from the individual cointegrating relationships. At first glance, it does not differ much from relationship (3), but Figure 8 shows a large discrepancy of the cumulative curves defined by (3) and (6). One can also use different individual relationships from Table 7b and obtain many versions of relationship (6).

A VECM with one cointegrating relation between the three variables and the maximum lag order 4 also defines a cumulative curve deviating from the observed one and that obtained using the integral approach. Figure 9 displays the two predicted curves, which diverge with time.

Figures 6 through 9 provide a visual evaluation of the deviation between the cumulative curves, which is often better than quantitative estimates. As expected, the coefficients obtained using the dynamic time series and such statistical methods as linear regression and VECM fail to accurately predict the evolution of price and unemployment level as function of labor force level. The agreement between the observed cumulative curves and those predicted using the integral approach is remarkable, however. Figure 10 demonstrates that fluctuations of the error terms in the cumulative relationships might be of lower amplitude than those in the dynamics relationships. Statistically, this means that the error terms in the dynamic relationships must be I(-1) not I(0). At the same time, the cumulative error terms are obviously characterized by a higher autocorrelation - the influence of every high amplitude fluctuation persists in time, but is completely compensated in several years by following counter-directed corrections. Therefore, in physical terms, the cumulative inflation (inflation index) is better described than the inflation itself, but standard statistical estimates of this fact are biased.

The gain obtained by the integral approach can be also demonstrated using quantitative estimates of the difference between measured and predicted values for dynamic and cumulative curves. Linear regression of the dynamic variable *GDPD* on *UE* and *dLF/LF* and regression of their cumulative versions give coefficients of regression lines and corresponding standard errors, *StErr*, i.e. the RMS deviation of the dependent variable from the straight lines. Another measure of the distance between the measured and predicted curves is defined as the root-mean square difference, *RMSD*, which is obtained using actual readings not regression line. This measure is important for cumulative curves obtained using regression coefficients for dynamic time series. Regression analysis can show an excellent correlation with a very low standard error for physically diverging curves.

Table 8 presents *StErr* and *RMSD* values resulted from the linear regressions and VECMs as estimated for the dynamic and cumulative curves. For each of the four relationships in the Table, we compare the row "Cumulative" to other five rows, where the estimates related to the coefficient listed in Tables 7a and 7b for the dynamic time series are given. The coefficients in the rows "Cumulative" have been obtained by the simplest procedure of a visual best fit between related cumulative curves, as shown in Figures 6 through 9. No minimization of standard error or maximization of overall correlation has been sought.

The values of *StErr* in the second column of Table 8 demonstrate that the integral approach provides practically the same accuracy as the regression and VECM carried out for the dynamic time series. The only marginal exception is the generalized (trivariate) relationship, where linear regression of the dynamic curves and the VECM with the maximum lag 3 give a lower value of 0.013 than 0.014 obtained by the integral approach. The *RMSD* values are larger or smaller than those of the *StErr* depending on synchronization of fluctuations. When measured and predicted curves fluctuate in sync, corresponding *RMSD* is smaller than *StErr*, and vice versa.

The *StErr* obtained using linear regression and VECM techniques demonstrate a principal similarity of the dynamic and cumulative time series. One can expect such a behavior when the link between the variables involved in the dynamic relationship is a true deterministic link, which also holds when the time series are differentiated or integrated. A problem for such a link are, however, measurement errors making statistical estimates of corresponding coefficients less reliable with increasing order of differentiation. Integrals are superior to differentials in suppression the measurement noise and increasing SNR for accurate estimates of the coefficients. The *RMSD* values in Table 8 are quite different for the dynamic and cumulative time series when corresponding coefficients are obtained using linear regression and the VECM representation. This discrepancy quantitatively confirms the visual effects observed in Figures 6 through 9. The integral approach provides very close estimates for the dynamic and cumulative time series in all four cases. Therefore the integral approach to the estimation of the coefficients in the linear lagged relationship between the inflation, unemployment and the change rate of the labor force is the most accurate for the dynamic and cumulative (level) variables.

## 5. Conclusion

The expected result of the above analysis consists in a formal statistical confirmation of the existence of a unique linear and lagged (four years for France) relationship between inflation, unemployment and labor force change rate for the period between 1973 and 2004. Hence, the three variables, being non-stationary I(1), are cointegrated in statistical sense; i.e. their residual time series in the VECM representation has been proved to be stationary. The absence of such a cointegration test was a weak point of Kitov (2007).

In a similar study of the relationship between GDP deflator and labor force change rate carried out for the USA (I.Kitov, O.Kitov & Dolinskaya, 2007), we have proved the existence of a cointegrating link during the period between 1960 and 2004. Monetary policy of the FRB differs from that implemented by the Bank de France in the absence of a fixed inflation target and any explicit limitation of monetary supply. As a result, the undisturbed link between inflation and labor force in the USA had to be replaced for France with the generalized relationship which also includes unemployment. The trivariate relationship, however, provides a very accurate prediction at a four-year

horizon – less than 1% for the entire period. The last twenty years, which are often called "Great Moderation" period, are characterized by even a lower uncertainty of the prediction at the same time horizon. The backyard of the French version of the "Great Moderation" is crowded by unemployed, however.

In our opinion, of the same importance for economics and econometrics is the introduction and development of an alternative technique for the analysis of true links between non-stationary variables – the integral approach or the usage of cumulative time series.

There are two drawbacks related to cointegration tests using the VAR representation that we would like to stress. The first one is associated with an increasing risk of rejection of a true equilibrium long-run relationship between non-stationary variables. The existence of a strict linear (or nonlinear) link between several non-stationary variables, a common case in physical sciences, implies that it works at any order of differentiation or integration in line with mathematical representation. Since measurement errors are an inevitable component of any actual link, their relative amplitude is of crucial importance, i.e. one can expect a good statistical inference for a large signal to noise ratio (SNR) and the increasing noise influence destroys the efficiency of statistical approach. When differenced, the variables involved in the true relationship loose their valid trend components which make a larger part of signal. Therefore, the differencing results in a significant decrease in the corresponding SNR and thus in deterioration of statistical inferences. Hence, the cointegration using VECM is a counterproductive method for revealing true relations between physical and economic variables. One always has to be very careful with the VECM approach and cointegration tests. It is very likely that many valid relationships between macroeconomic variables reside in garbage bins as rejected by cointegration tests and waiting for a rediscovery.

We have proposed an approach, which results in increasing SNR and involves additional and accurate information on measured values – the usage of cumulative curves. By definition, such an approach works only in the case of the existence of a true and strict link between measured variables. The existence of such strict links in economics, not only in physical sciences, is demonstrated in our papers devoted to the modeling of such macroeconomic variable as inflation as a linear and lagged function of unemployment rate and labor force change rate. Fortunately, in the USA and France measurement noise characterizing the variables is small enough for a VECM to provide an adequate representation for cointegration tests, which do reject the null hypothesis of the presence of unit roots in error terms. For different countries, however, the existence of the link between dynamic and cumulative variables does not guarantee the rejection of the null hypothesis due to such specific properties of relevant measurements as piece-wise systematic errors.

The second drawback is defined by measurement noise properties of actual macroeconomic variables. The VECM representation assumes that error term is i.i.d. with zero mean and finite variance. This is not the case, however, in those economic time series which are obtained using population controls (I.Kitov, O.Kitov & Dolinskaya, 2007). Even a very low amplitude noise term with piece-wise systematic error induced by measuring procedure results in the acceptance of the unit root presence in corresponding residual series. So, two variables differing by few hundredths of their amplitude, i.e. practically indistinguishable by visual inspection, are considered as not cointegrated according to such cointegration tests.

For many time series, such as unemployment, there is no opportunity to carry out a re-estimation back in the past according to modern definitions and procedures. Therefore, the systematic errors induced by numerous changes in enumeration procedures and definitions are not removable from the series and standard cointegration tests are probably to give wrong result if not mixed with some quasi-white or pure white noise associated with random measurement errors. For cointegration tests to be successful in terms of rejection of spurious regression and acceptance of a true relationship, one has to retain the random noise of significant amplitude, i.e. to deny improvements in measuring procedures. This contradicts fundamental principles of the scientific approach and is thus unacceptable.

The integral approach provides accurate estimates for coefficients of the link between dynamic and cumulative time series and does not depend on "poor" statistical properties of the noise term in the annual readings.

**Acknowledgments**

The authors are grateful to Dr. Wayne Richardson for his constant interest in this study, help and assistance, and fruitful discussions. The manuscript was greatly improved by his critical review.

**Tables**

Table 1. Descriptive statistics for the original time series and their first differences

| Variable | GDPD | CPI | dLF/LF | UE | dGDPD | dCPI | d(dLF/LF) | dUE |
|---|---|---|---|---|---|---|---|---|
| mean | 5.3E-2 | 5.3E-2 | 6.6E-3 | 6.4E-2 | -1.4E-3 | 4.9E-5 | -5.7E-5 | 1.9E-3 |
| st. dev. | 4.2E-2 | 4.0E-2 | 4.1E-3 | 4.0E-2 | 1.2E-2 | 2.7E-2 | 4.3E-3 | 5.7E-3 |
| skewness | 4.6E-1 | 9.9E-1 | -1.8E-1 | -1.2E-2 | 3.1E-1 | 1.2E+0 | 1.6E-1 | -9.7E-1 |
| kurtosis | 1.6E+0 | 2.8E+0 | 2.8E+0 | 1.4E+0 | 3.6E+0 | 1.2E+1 | 2.3E+0 | 5.1E+0 |

Table 2. Unit root tests for GDP deflator, CPI, unemployment, and labor force change rate

| Variable | ADF [lag] | | DF_GLS [lag] | |
| --- | --- | --- | --- | --- |
| | 0 | 1 | 1 | 2 |
| GDP deflator (1972-2004) | -0.63 (-3.70) | -1.12 (-3.71) | -1.61 (-3.77) | -1.70 (-3.77) |
| CPI (1957-2004) | -2.44 (-3.60) | -2.11 (-3.61) | -1.72 (-3.77) | -1.64 (-3.77) |
| UE (1958-2004) | -1.06 (-3.60) | -1.12 (-3.61) | -2.29 (-3.77) | -1.93 (-3.77) |
| dLF/LF (1957-2004) | -4.09* (-3.60) | -3.22 (-3.61) | -3.43 (-3.77) | -2.62 (-3.77) |

\* - rejection of the null hypothesis of the unit root presence

Table 3. Unit root tests for the first difference of GDP deflator, CPI, unemployment, and labor force change rate in Table 1

| First difference variable | ADF [lag] | | | | DF_GLS [lag] | | | |
|---|---|---|---|---|---|---|---|---|
| | 0 | 1 | 2 | 3 | 1 | 2 | 3 | 4 |
| GDPD (trend) (1972-2004) | -3.85 (-4.32) | -3.32 (-4.33) | -4.41 (-4.33) | -3.57 (-4.33) | -2.87 (-3.77) | -2.94 (-3.70) | -2.00 (-3.77) | -2.23 (-3.77) |
| GDPD (constant) (1972-2004) | -3.90* (-3.70) | -3.53 (-3.71) | -4.57* (-3.72) | -3.50 (-3.72) | -2.69* (-2.65) | -2.69* (-2.65) | -1.83 (-2.65) | -1.95 (-2.65) |
| CPI (trend) (1957-2004) | -8.09* (-4.18) | -7.91* (-4.19) | -5.24* (-4.20) | -3.65 (-4.21) | -5.02* (-3.77) | -5.17* (-3.77) | -3.65 (-3.77) | -2.85 (-3.77) |
| CPI (constant) (1957-2004) | -8.12* (-3.60) | -8.13* (-3.61) | -5.24* (-3.61) | -3.63* (-3.62) | -4.37* (-2.63) | -4.42* (-2.63) | -3.00* (-2.63) | -2.27 (-2.63) |
| UE (trend) (1958-2004) | -4.06 (-4.19) | -3.71 (-4.20) | -3.25 (-4.21) | -3.44 (-4.21) | -2.29 (-3.77) | -1.93 (-3.77) | -1.89 (-1.87) | -1.54 (-3.77) |
| UE (constant) (1958-2004) | -5.65* (-3.60) | -3.91* (-3.61) | -3.09 (-3.62) | -3.17 (-3.63) | -3.60* (-2.63) | -3.15* (-2.63) | -3.31* (-2.63) | -2.64* (-2.63) |
| dLF/LF (trend) (1958-2004) | -10.02* (-4.18) | -8.35* (-4.18) | -5.46* (-4.20) | -3.61 (-4.21) | -7.60* (-3.77) | -4.57* (-3.77) | -3.02 (-3.77) | -2.84 (-3.77) |
| dLF/LF (constant) (1958-2004) | -10.10* (-3.60) | -8.36* (-3.61) | -5.33* (-3.61) | -3.42 (-3.62) | -6.99* (-2.63) | -4.05* (-2.63) | -2.58 (-2.63) | -2.37 (-2.63) |

*- rejection of the null hypothesis of the unit root presence

Table 4. Unit root tests for the residuals of relationship (3)

| Variable | ADF [lag] | | | | DF_GLS [lag] | | | |
|---|---|---|---|---|---|---|---|---|
| | 0 | 1 | 2 | 3 | 1 | 2 | 3 | 4 |
| diff1 | -4.65* (-3.70) | -4.55* (-3.70) | -3.61 (-3.71) | -3.47 (-3.72) | -1.39 (-2.65) | -1.30 (-2.65) | -1.28 (-2.65) | -1.07 (-2.65) |
| diff2 | -3.38 (-3.70) | -4.82* (-3.70) | -3.83* (-3.71) | -3.83* (-3.72) | -3.04* (-2.65) | -3.92* (-2.65) | -3.65* (-2.65) | -3.51* (-2.65) |
| diff3 | -3.10 (-3.70) | -3.74* (-3.70) | -3.91* (-3.71) | -3.30 (-3.72) | -3.48* (-2.65) | -3.88* (-2.65) | -3.34* (-2.65) | -5.10* (-2.65) |
| diff4 | -3.19 (-3.70) | -4.37* (-3.70) | -4.23* (-3.71) | -3.50 (-3.72) | -3.76* (-2.65) | -2.86* (-2.65) | -2.74* (-2.65) | -3.11* (-2.65) |

*- rejection of the null hypothesis of the unit root presence

Table 5a. Specifications tests for GDP deflator as a function of UE and dLF/LF for the period between 1971 and 2004

| Predictor | Hettest [1] Pr>chi2 | Ramsey [2] test Pr>F | LM for ARCH [3] Pr>chi2 | Breusch-Godfrey LM [4] Pr>chi2 | DW [5] d-stat | $R^2$ | RMS(F)E | Cons [cons] Pr>|t| |
|---|---|---|---|---|---|---|---|---|
| *UE+dLF/LF* | 0.014 | 0.07 | 0.016 | 0.0017 | 0.82 | 0.85 | 0.017 | 0.10 [0.01] 0.000 |
| *predicted* | 0.04 | 0.098 | 0.10 | 0.02 | 0.99 | 0.84 | 0.017 | 0.0054 [0.0045] 0.24 |
| *MA(2)* | 0.04 | 0.09 | 0.93 | 0.006 | 1.06 | 0.9 | 0.013 | 0.002 [0.004] 0.55 |
| *MA(3)* | 0.31 | 0.004 | 0.68 | 0.002 | 0.94 | 0.91 | 0.012 | 0.003 [0.004] 0.93 |

1) $H_0$ - constant variance; 2) $H_0$ - no omitted variables; 3) $H_0$ - no ARCH effect; 4) $H_0$ - no serial correlation; 5) $H_0$ - no serial correlation

Table 5b. Specifications tests for GDP deflator as a function of dLF/LF for the period between 1971 and 1999

| GDP deflator | hettest Pr>chi2 | Ramsey test Pr>F | LM for ARCH Pr>chi2 | Breusch-Godfrey LM Pr>chi2 | DW d-stat | $R^2$ | RMS(F)E | cons [cons] Pr>\|t\| |
|---|---|---|---|---|---|---|---|---|
| *predicted* | 0.52 | 0.03 | 0.81 | 0.20 | 1.52 | 0.73 | 0.022 | 0.013 [0.006] 0.04 |
| *MA(2)* | 0.90 | 0.08 | 0.40 | 0.08 | 1.37 | 0.87 | 0.016 | 0.005 [0.005] 0.34 |
| *MA(3)* | 0.11 | 0.02 | 0.21 | 0.05 | 1.19 | 0.93 | 0.012 | 0.0008 [0.004] 0.83 |

Table 5c. Specifications tests for CPI as a function of dLF/LF for the period between 1967 and 1999

| CPI | hettest Pr>chi2 | Ramsey test Pr>F | LM for ARCH Pr>chi2 | Breusch-Godfrey LM Pr>chi2 | DW d-stat | $R^2$ | RMS(F)E | cons [cons] Pr>|t| |
|---|---|---|---|---|---|---|---|---|
| *Predicted* | 0.82 | 0.67 | 0.83 | 0.001 | 1.18 | 0.43 | 0.031 | 0.029 [0.008] 0.001 |
| *MA(2)* | 0.39 | 0.51 | 0.49 | 0.09 | 1.44 | 0.71 | 0.022 | 0.007 [0.007] 0.29 |
| *MA(3)* | 0.86 | 0.05 | 0.62 | 0.02 | 1.16 | 0.83 | 0.017 | 0.005 [0.005] 0.44 |

Table 5d. Specifications tests for UE as a function of dLF/LF for the period between 1971 and 1995

| UE | hettest Pr>chi2 | Ramsey test Pr>F | LM for ARCH Pr>chi2 | Breusch-Godfrey LM Pr>chi2 | DW d-stat | $R^2$ | RMS(F)E | cons [cons] Pr>|t| |
|---|---|---|---|---|---|---|---|---|
| *Predicted* | 0.03 | 0.02 | 0.89 | 0.10 | 1.36 | 0.71 | 0.017 | 0.021 [0.008] 0.013 |

Table 6a. Cointegration rank test. GDP deflator vs. UE and dLF/LF during the period between 1973 and 2004.

| Predictor | Trend specification | Rank | Lag | LL | Eigenvalue | Trace statistics | 5% critical value |
|---|---|---|---|---|---|---|---|
| UE+dLF/LF | constant | 1 | 1 | 352.1 | 0.682 | 9.91* | 15.41 |
| | constant | 1 | 2 | 357.8 | 0.625 | 14.63* | 15.41 |
| | constant | 0 | 3 | 349.6 | . | 23.51* | 29.68 |
| | constant | 1 | 4 | 360.0 | 0.605 | 3.83* | 15.41 |
| | rconstant | 1 | 1 | 349.5 | 0.694 | 15.09* | 19.96 |
| | rconstant | 2 | 2 | 361.2 | 0.353 | 7.73* | 9.42 |
| | rconstant | 0 | 3 | 346.4 | . | 29.92* | 34.91 |
| | rconstant | 1 | 4 | 357.1 | 0.614 | 9.69* | 19.96 |
| | none | 1 | 1 | 349.5 | 0.694 | 6.13* | 12.53 |
| | none | 1 | 2 | 354.7 | 0.627 | 8.91* | 12.53 |
| | none | 0 | 3 | 346.4 | . | 14.83* | 24.31 |
| | none | 0 | 4 | 343.8 | . | 16.48* | 24.31 |
| predicted | constant | 1 | 1 | 185.3 | 0.415 | 0.62* | 3.76 |
| | constant | 0 | 2 | 180.8 | . | 12.63* | 15.41 |
| | constant | 1 | 3 | 183.8 | 0.359 | 3.68* | 3.76 |
| | constant | 1 | 4 | 186.6 | 0.480 | 2.89* | 3.76 |
| | rconstant | 0 | 1 | 176.4 | . | 18.36* | 19.96 |
| | rconstant | 0 | 2 | 179.5 | . | 15.24* | 19.96 |
| | rconstant | 1 | 3 | 181.5 | 0.366 | 8.28* | 9.42 |
| | rconstant | 1 | 4 | 185.7 | 0.485 | 4.70* | 9.42 |
| MA(3) | constant | 1 | 1 | 213.2 | 0.407 | 0.42* | 3.76 |
| | constant | 0 | 2 | 211.5 | . | 12.43* | 15.41 |
| | constant | 2 | 3 | 215.5 | 0.126 | | |
| | constant | 1 | 4 | 210.9 | 0.470 | 2.48* | 3.76 |
| | rconstant | 1 | 1 | 212.1 | 0.458 | 2.70* | 9.42 |
| | rconstant | 0 | 2 | 209.7 | . | 15.90* | 19.96 |
| | rconstant | 1 | 3 | 212.1 | 0.455 | 6.66* | 9.42 |
| | rconstant | 1 | 4 | 210.5 | 0.480 | 3.44* | 9.42 |

Table 6b. Cointegration rank test. Observed vs. predicted GDP deflator for the period between 1971 and 1999.

| Predictor | Trend specification | Rank | Lag | LL | Eigenvalue | Trace statistics | 5% critical value |
|---|---|---|---|---|---|---|---|
| dLF/LF | constant | 1 | 1 | 227.5 | 0.579 | 0.02* | 3.76 |
| | constant | 1 | 2 | 231.8 | 0.726 | 0.46* | 3.76 |
| | constant | 0 | 3 | 219.1 | . | 11.65* | 15.41 |
| | constant | 0 | 4 | 216.8 | . | 12.25* | 15.41 |
| | rconstant | 1 | 1 | 227.1 | 0.580 | 0.94* | 9.42 |
| | rconstant | 1 | 2 | 231.4 | 0.726 | 1.28* | 9.42 |
| | rconstant | 0 | 3 | 218.3 | . | 13.43* | 19.96 |
| | rconstant | 0 | 4 | 213.9 | . | 17.96* | 19.96 |
| MA(2) | constant | 1 | 1 | 160.5 | 0.447 | 0.01* | 3.76 |
| | constant | 1 | 2 | 162.8 | 0.647 | 0.44* | 3.76 |
| | constant | 0 | 3 | 155.9 | . | 10.89* | 15.41 |
| | constant | 0 | 4 | 157.6 | . | 14.01* | 15.41 |
| | rconstant | 0 | 1 | 151.7 | . | 17.59* | 19.96 |
| | rconstant | 1 | 2 | 162.4 | 0.647 | 1.24* | 9.42 |
| | rconstant | 0 | 3 | 155.1 | . | 12.63* | 19.96 |
| | rconstant | 0 | 4 | 154.7 | . | 19.68* | 19.96 |

Table 6c. Cointegration rank test. Observed vs. predicted CPI inflation for the period between 1967 and 1999.

| Predictor | Trend specification | Rank | Lag | LL | Eigenvalue | Trace statistics | 5% critical value |
|---|---|---|---|---|---|---|---|
| dLF/LF | constant | 1 | 1 | 233.4 | 0.697 | 0.89* | 3.76 |
| | constant | 1 | 2 | 233.2 | 0.696 | 1.39* | 3.76 |
| | constant | 1 | 3 | 226.2 | 0.419 | 0.97* | 3.76 |
| | constant | 1 | 4 | 219.9 | 0.417 | 0.70* | 3.76 |
| | rconstant | 1 | 1 | 233.4 | 0.697 | 0.98* | 9.42 |
| | rconstant | 1 | 2 | 233.2 | 0.696 | 1.53* | 9.42 |
| | rconstant | 0 | 3 | 217.9 | . | 17.72* | 19.96 |
| | rconstant | 0 | 4 | 212.0 | . | 16.70* | 19.96 |
| MA(2) | constant | 1 | 1 | 159.7 | 0.500 | 0.96* | 3.76 |
| | constant | 1 | 2 | 161.2 | 0.617 | | 3.76 |
| | constant | 0 | 3 | 153.1 | . | 12.15* | 15.41 |
| | constant | 0 | 4 | 148.4 | . | 14.99* | 15.41 |
| | rconstant | 1 | 1 | 159.6 | 0.500 | 1.01* | 9.42 |
| | rconstant | 1 | 2 | 161.1 | 0.618 | 1.64* | 9.42 |
| | rconstant | 0 | 3 | 152.8 | . | 12.77* | 19.96 |
| | rconstant | 0 | 4 | 148.1 | . | 15.60* | 19.9 |

Table 6d. Cointegration rank test. Observed vs. predicted unemployment for the period between 1971 and 1995.

| Predictor | Trend specification | Rank | Lag | LL | Eigenvalue | Trace statistics | 5% critical value |
|---|---|---|---|---|---|---|---|
| dLF/LF | constant | 1 | 1 | 201.7 | 0.671 | 0.88* | 3.76 |
| | constant | 1 | 2 | 199.2 | 0.634 | 3.06* | 3.76 |
| | constant | 1 | 3 | 192.7 | 0.474 | 2.92* | 3.76 |
| | constant | 0 | 4 | 178.2 | . | 14.96* | 15.41 |
| | rconstant | 1 | 1 | 197.7 | 0.676 | 8.81* | 9.42 |
| | rconstant | 1 | 2 | 197.9 | 0.634 | 5.65* | 9.42 |
| | rconstant | 1 | 3 | 190.9 | 0.475 | 6.57* | 9.42 |
| | rconstant | 0 | 4 | 176.4 | . | 18.56* | 19.96 |

Table 7a. Coefficients in the cointegrating relation(s) as a function of the maximum lag order in VEC representation.

| | GDPD vs. UE and dLF/LF (rank 1) | | | | | |
|---|---|---|---|---|---|---|
| Lag | Slope UE | St. Err. | Slope dLF | St. Err. | Intercept | RMSE |
| 1 | -0.497 | 0.135 | 11.48 | 1.28 | 0.0037 | 0.0123 |
| 2 | -0.591 | 0.114 | 11.29 | 1.38 | 0.0186 | 0.0097 |
| 3 | -1.050 | 0.127 | 3.12 | 1.74 | 0.106 | 0.0085 |
| 4 | -0.957 | 0.095 | 5.11 | 1.33 | 0.0885 | 0.0082 |
| | GDPD vs. dLF/LF and UE vs. dLF/LF (rank 2) | | | | | |
| | GDPD Slope | St. Err. | UE Slope | St. Err. | Intercept | RMSE |
| 1 | 18.90 | 2.07 | -14.92 | 2.31 | 0.0926 | 0.0113 |
| 2 | 20.83 | 2.07 | -16.15 | 2.40 | 0.0962 | 0.0092 |
| 3 | 48.25 | 16.80 | -43.00 | 16.11 | 0.0935 | 0.0084 |
| 4 | 64.91 | 32.00 | -62.48 | 32.88 | 0.1076 | 0.0082 |

Table 7b. Coefficients in the cointegrating relation as a function of the maximum lag order in VECM representation

| | UE vs. dLF/LF | | | |
|---|---|---|---|---|
| Lag | Slope | StErr | Intercept | RMSE |
| 1 | -12.62 | 8.47 | 0.174 | 0.0057 |
| 2 | -11.97 | 1.20 | 0.157 | 0.0054 |
| 3 | -10.75 | 8.86 | 0.157 | 0.0055 |
| 4 | -10.91 | 1.20 | 0.156 | 0.0056 |
| | GDPD vs. dLF/LF | | | |
| | Slope | StErr | Intercept | RMSE |
| 1 | 17.97 | 1.51 | -0.086 | 0.013 |
| 2 | 17.45 | 0.85 | -0.063 | 0.012 |
| 3 | 17.66 | 1.11 | -0.064 | 0.012 |
| 4 | 17.76 | 1.00 | -0.0508 | 0.011 |

Table 8. Comparison of standard errors obtained from regressions and RMS differences (RMSD) for dynamic and cumulative curves (see text for details)

|  | Dynamic | | Cumulative | |
|---|---|---|---|---|
|  | StErr | RMSD | StErr | RMSD |
| UE vs. dLF | | | | |
| Cumulative | 0.024 | 0.035 | 0.029 | 0.042 |
| Linear regression | 0.024 | 0.023 | 0.057 | 0.129 |
| VECM Lag 1 | 0.024 | 0.035 | 0.021 | 0.198 |
| 2 | 0.024 | 0.032 | 0.024 | 0.043 |
| 3 | 0.024 | 0.030 | 0.022 | 0.161 |
| 4 | 0.024 | 0.030 | 0.021 | 0.130 |
| GDP vs. dLF/LF | | | | |
| Cumulative | 0.031 | 0.042 | 0.034 | 0.036 |
| Linear regression | 0.031 | 0.029 | 0.098 | 0.192 |
| VECM Lag 1 | 0.031 | 0.046 | 0.061 | 0.222 |
| 2 | 0.031 | 0.062 | 0.036 | 0.102 |
| 3 | 0.031 | 0.044 | 0.036 | 0.114 |
| 4 | 0.031 | 0.049 | 0.053 | 0.347 |
| GDPD vs. dLF/LF and UE (rank 1) | | | | |
| Cumulative | 0.014 | 0.014 | 0.017 | 0.017 |
| Linear regression | 0.013 | 0.012 | 0.025 | 0.024 |
| VECM Lag 1 | 0.035 | 0.033 | 0.030 | 0.098 |
| 2 | 0.026 | 0.032 | 0.028 | 0.058 |
| 3 | 0.013 | 0.013 | 0.020 | 0.054 |
| 4 | 0.015 | 0.016 | 0.017 | 0.026 |
|  | | | | |
| GDPD vs. dLF/LF and UE vs. dLF/LF (rank 2) | | | | |
| Cumulative | 0.031 | 0.042 | 0.034 | 0.036 |
| VECM Lag 1 | 0.031 | 0.046 | 0.061 | 0.222 |
| 2 | 0.031 | 0.062 | 0.036 | 0.102 |
| 3 | 0.031 | 0.044 | 0.036 | 0.114 |
| 4 | 0.031 | 0.049 | 0.053 | 0.347 |

**Figures**

a)

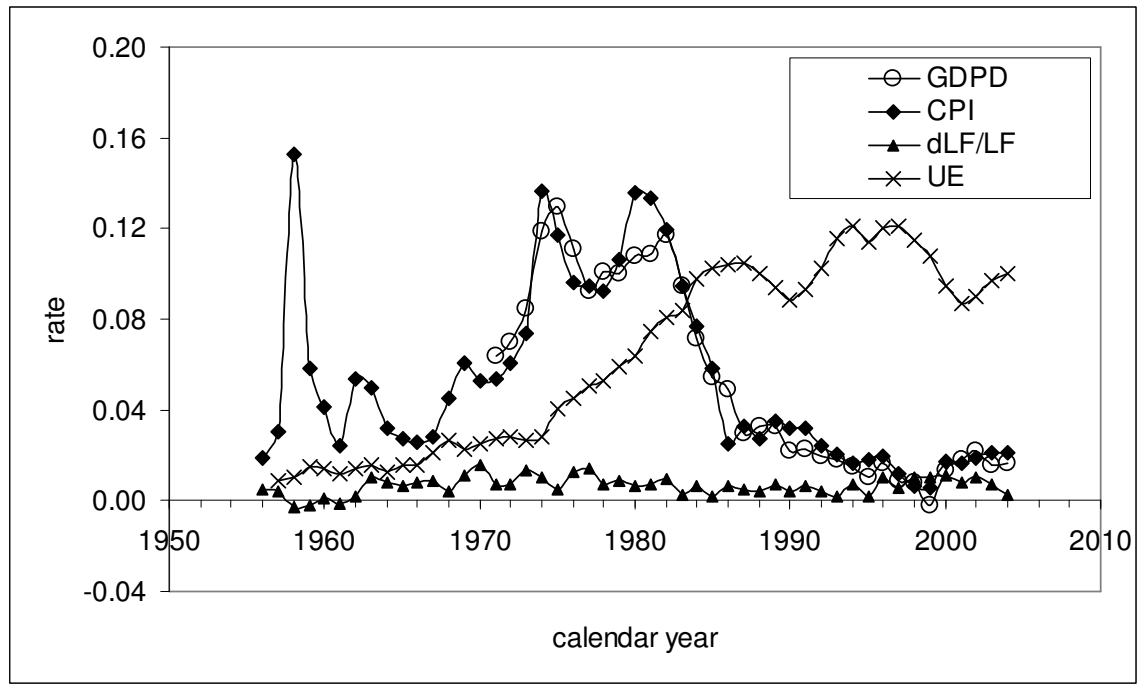

b)

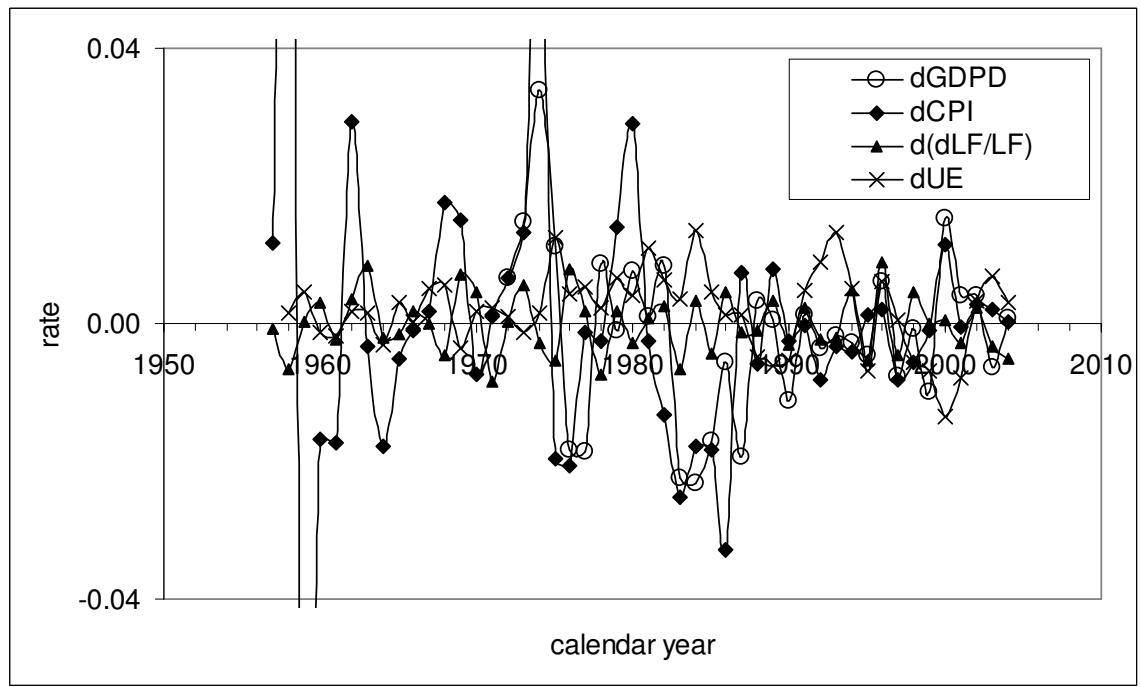

Figure 1. Measured time series in France: GDP deflator, CPI, unemployment, and labor force change rate – a); and their first differences – b).

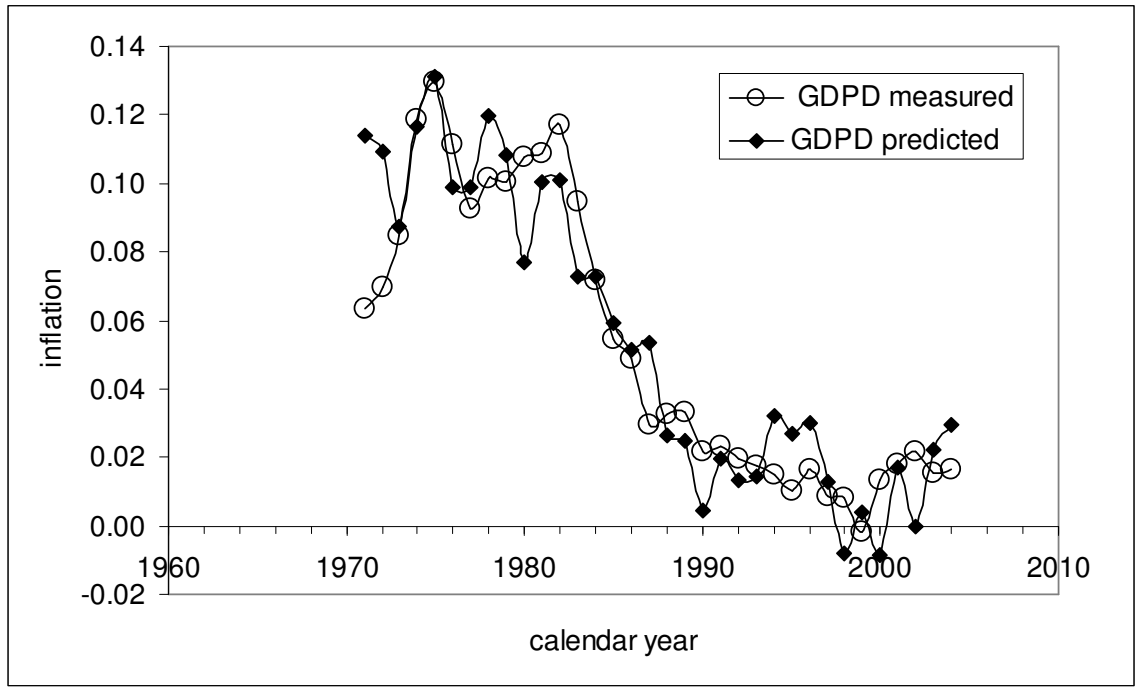

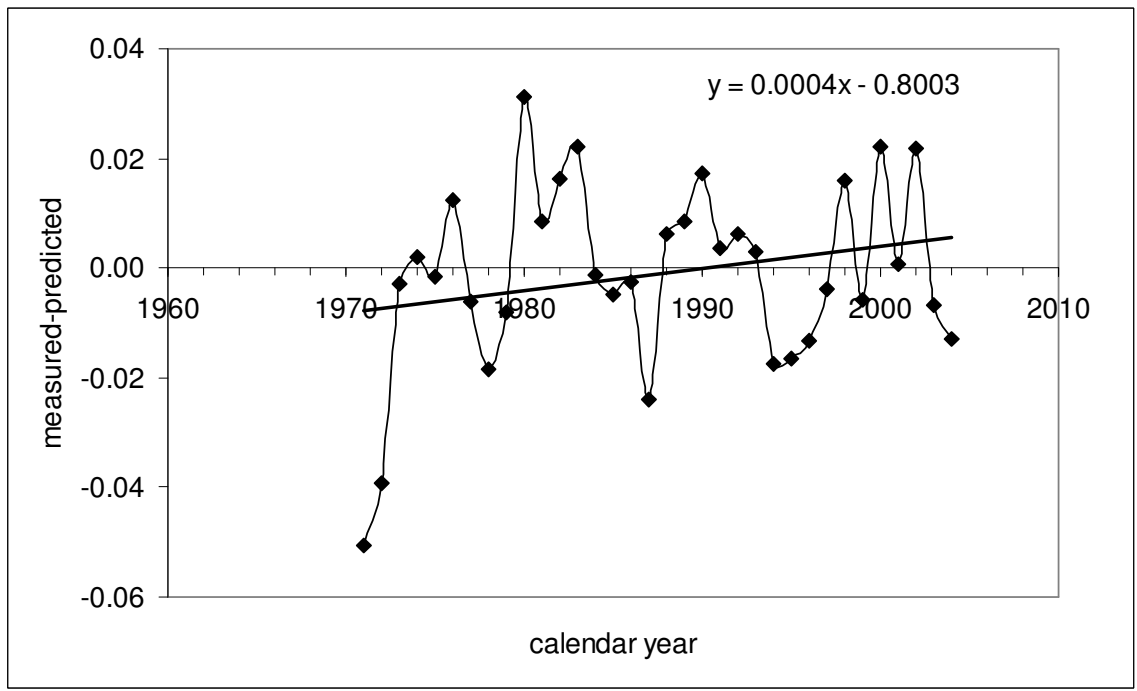

Figure 2. Measured and predicted GDP deflator - a), and their difference - b). The predicted time series is based on the measured labor force change rate and unemployment. Notice a strong side effect near the start of the series. Due to the limited length of the series, the side effect may severely alter statistical inferences.

a)

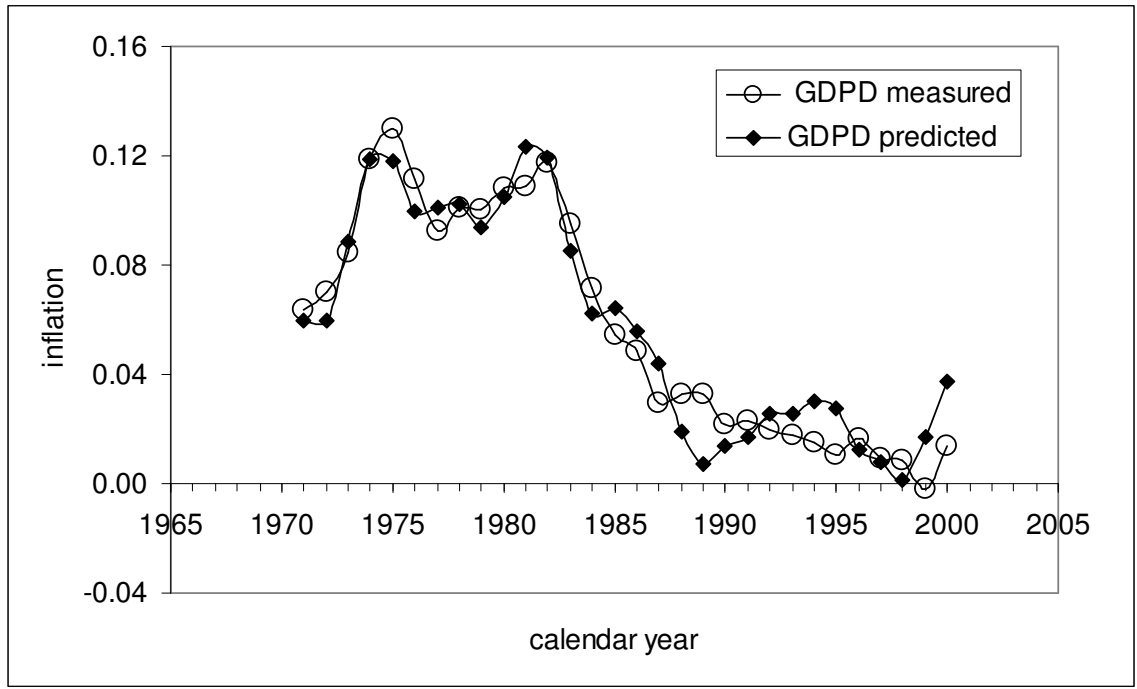

b)

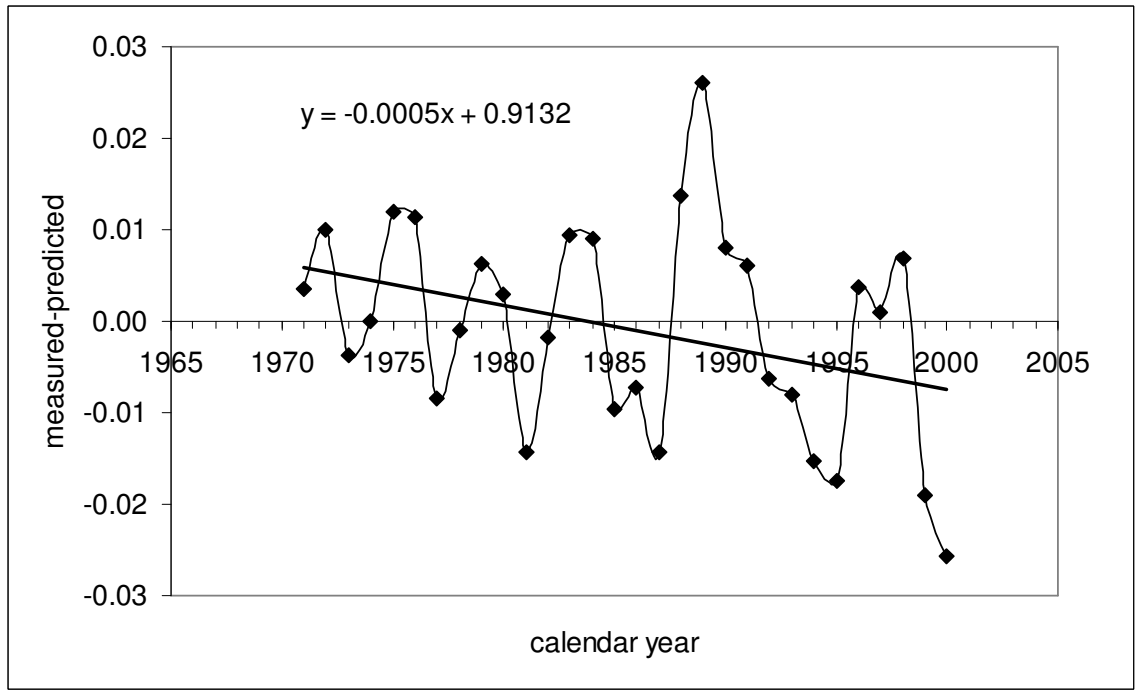

Figure 3. Measured and predicted GDP deflator - a), and their difference - b). The predicted time series between 1971 and 1999 is based on the measured labor force change rate only and is represented by *MA(3)* of the original predicted series. Notice the autocorrelation effect introduced by the moving average. The effect results in a biased statistical inference.

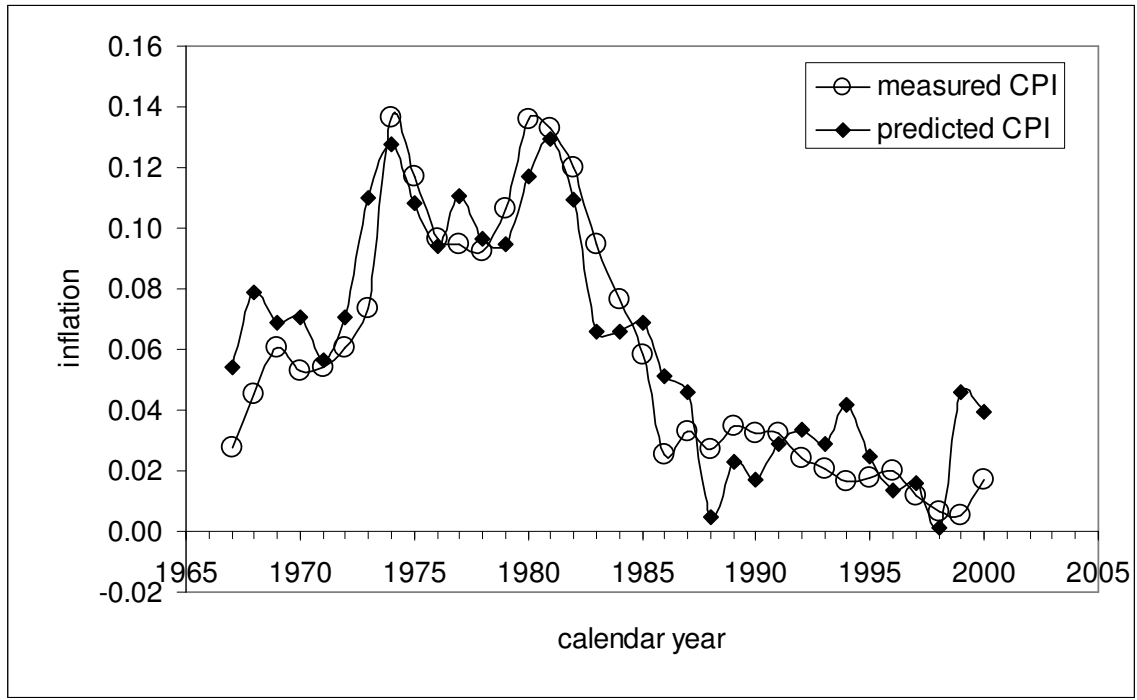
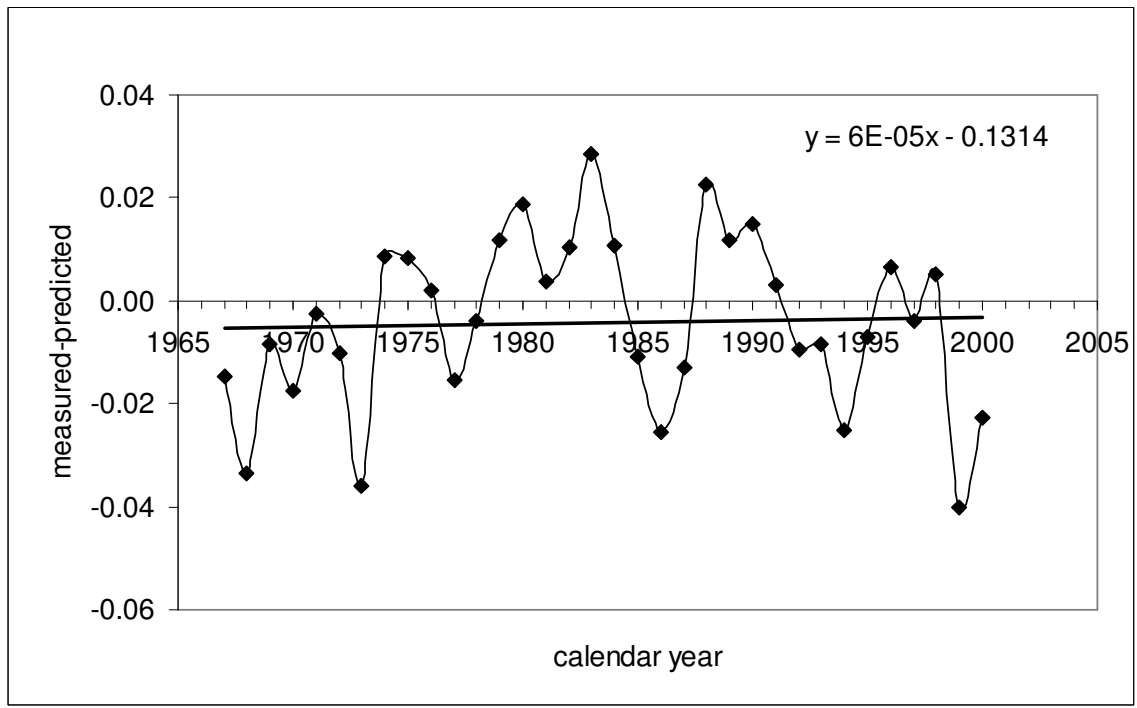

Figure 4. Measured and predicted CPI - a), and their difference - b). The predicted time series between 1967 and 1999 is based on the measured labor force change rate only and is represented by *MA(3)* of the original predicted series.

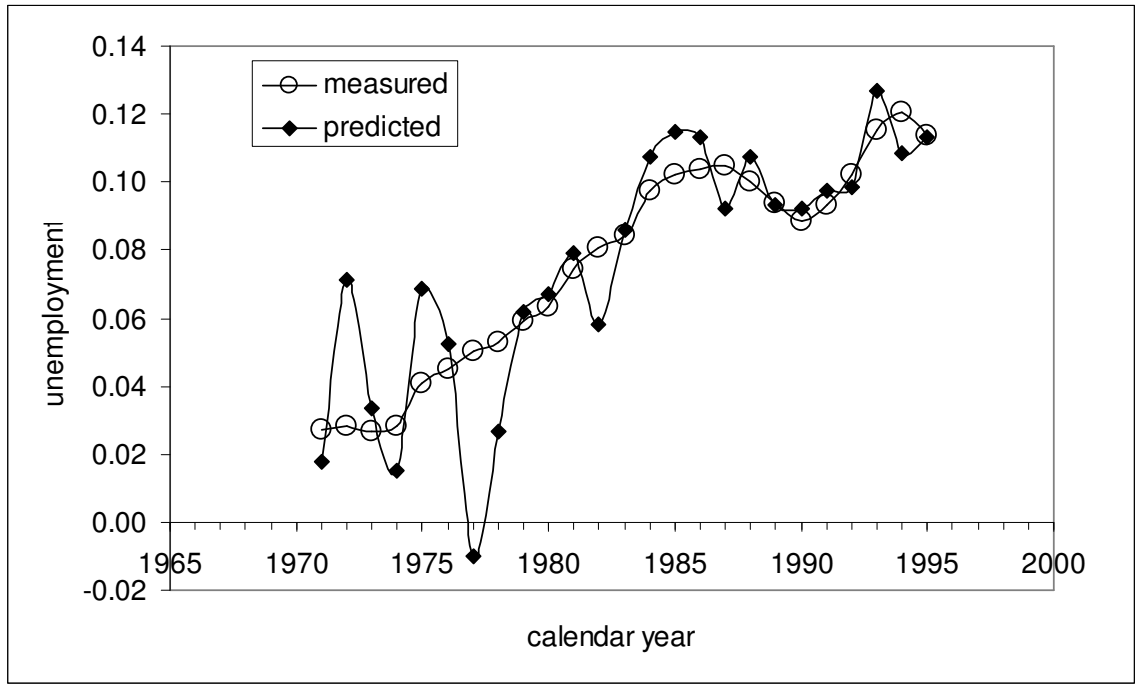

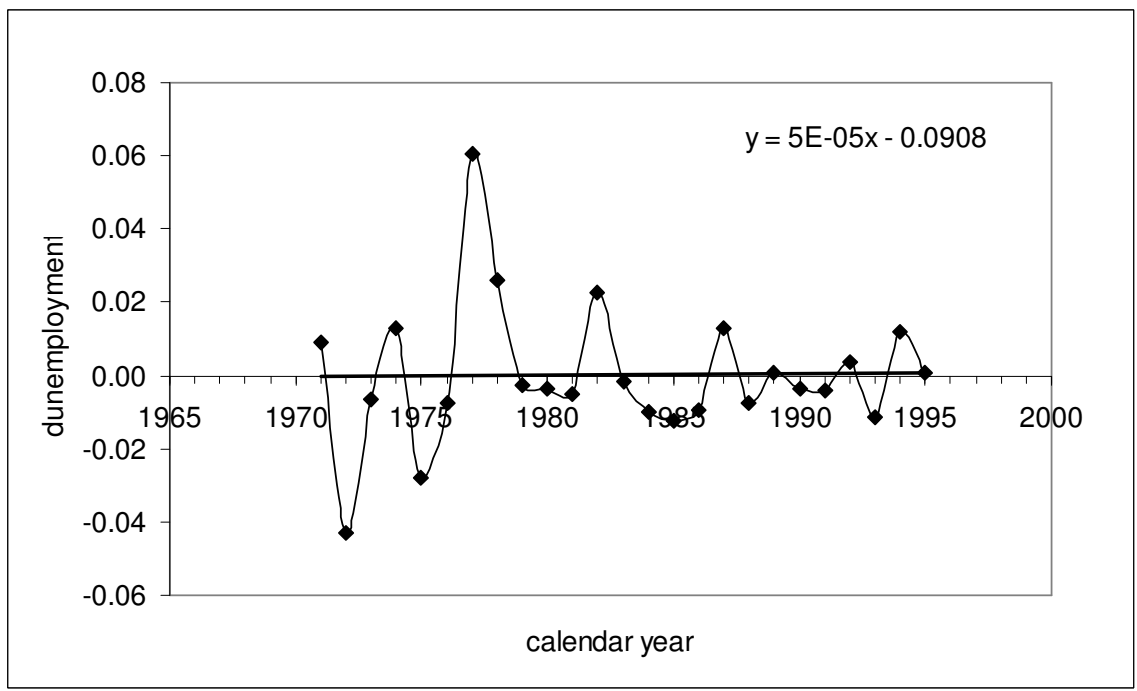

Figure 5. Measured and predicted unemployment - a); and their difference - b). The predicted time series between 1971 and 1995 is based on the measured labor force change rate only.

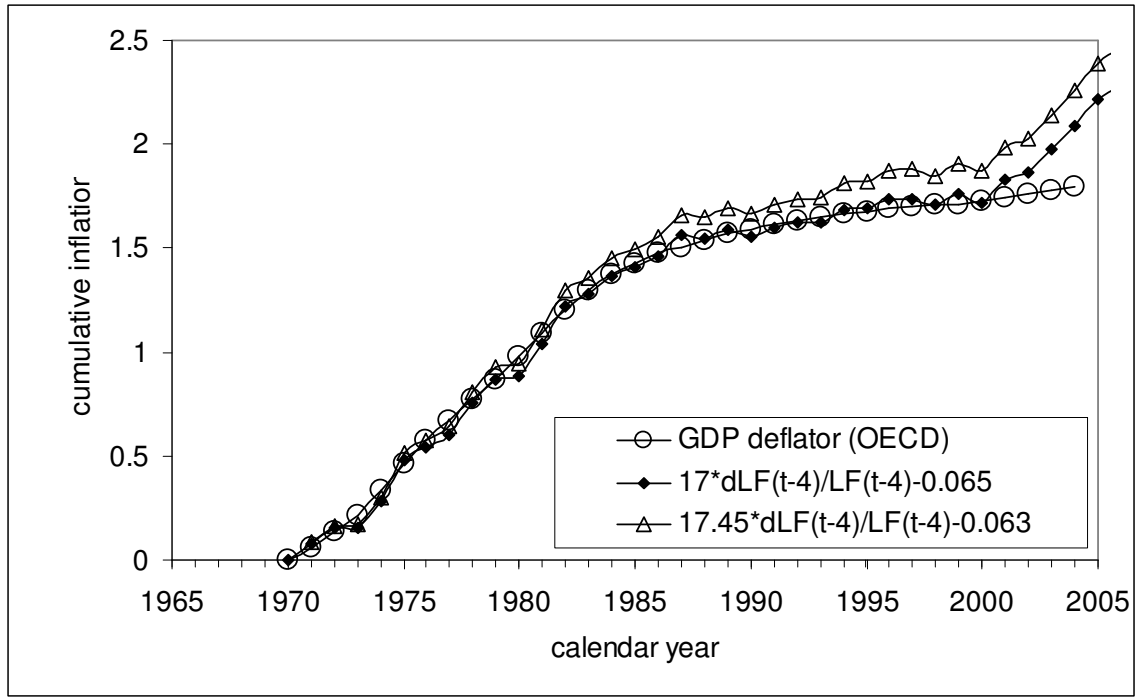

Figure 6. Cumulative curves for measured and predicted GDP deflator. The latter curves are obtained using the cumulative (solid diamonds) and dynamic (open triangles) time series. Open circles show the observed cumulative curve.

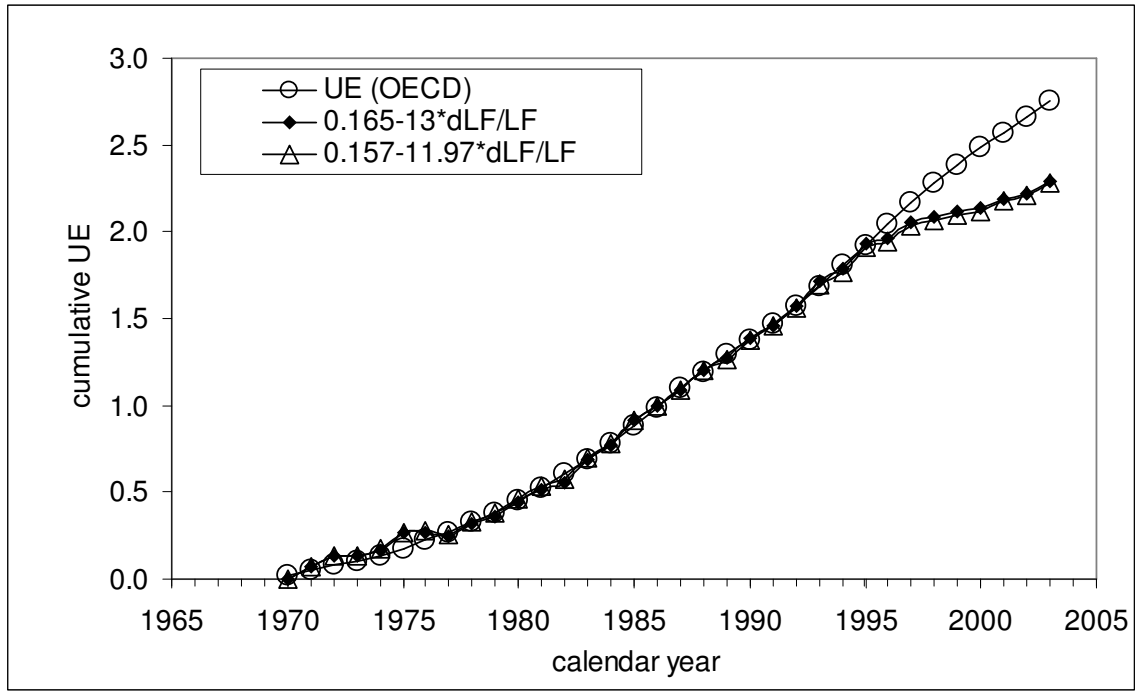

Figure 7. Cumulative curves for measure and predicted unemployment. The latter curves are obtained using the cumulative (solid diamonds) and dynamic (open triangles) time series. Open circles show the observed cumulative curve.

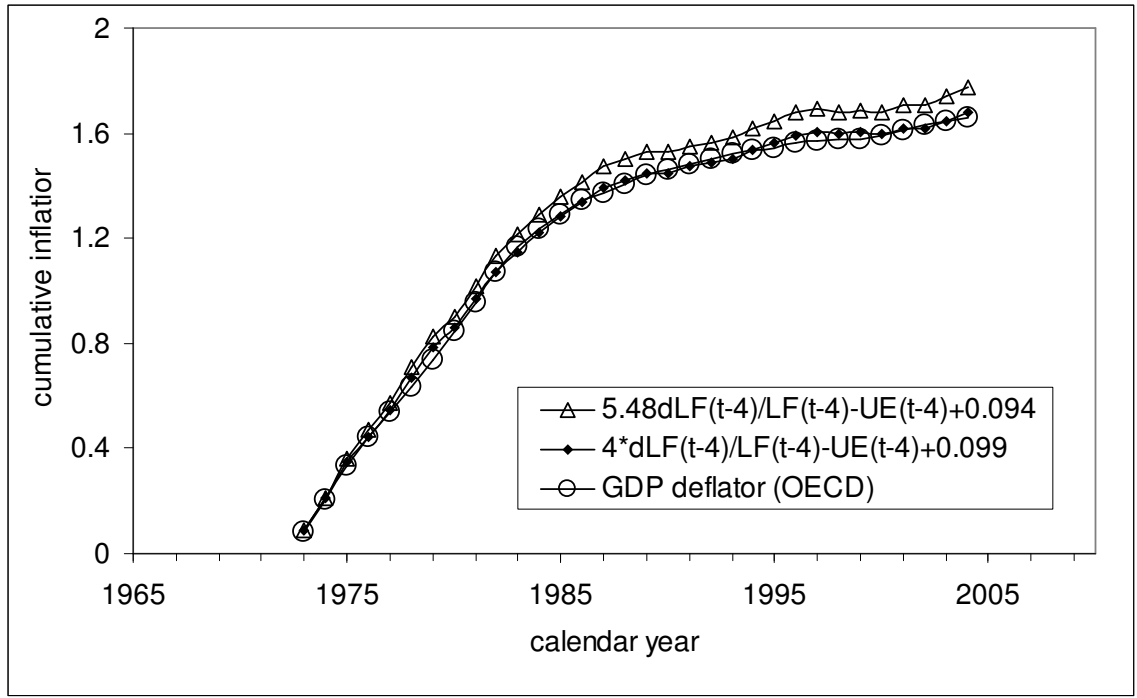

Figure 8. Cumulative curves of for measured and predicted inflation. The latter curves are obtained using cumulative (solid diamonds) and dynamic (open triangles) time series. Open circles show the observed cumulative curve.

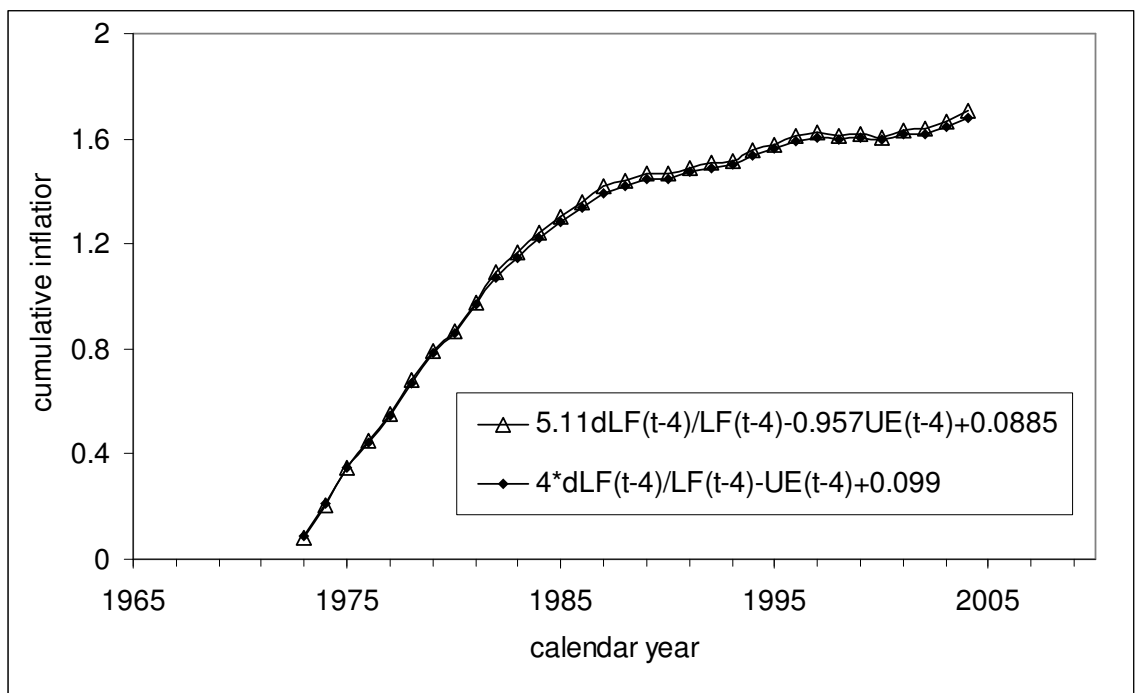

Figure 9. Cumulative curves for predicted inflation obtained using coefficients estimated from the cumulative (solid diamonds) and dynamic (open triangles) time series. Corresponding relationships are shown in the lower right corner.

a)

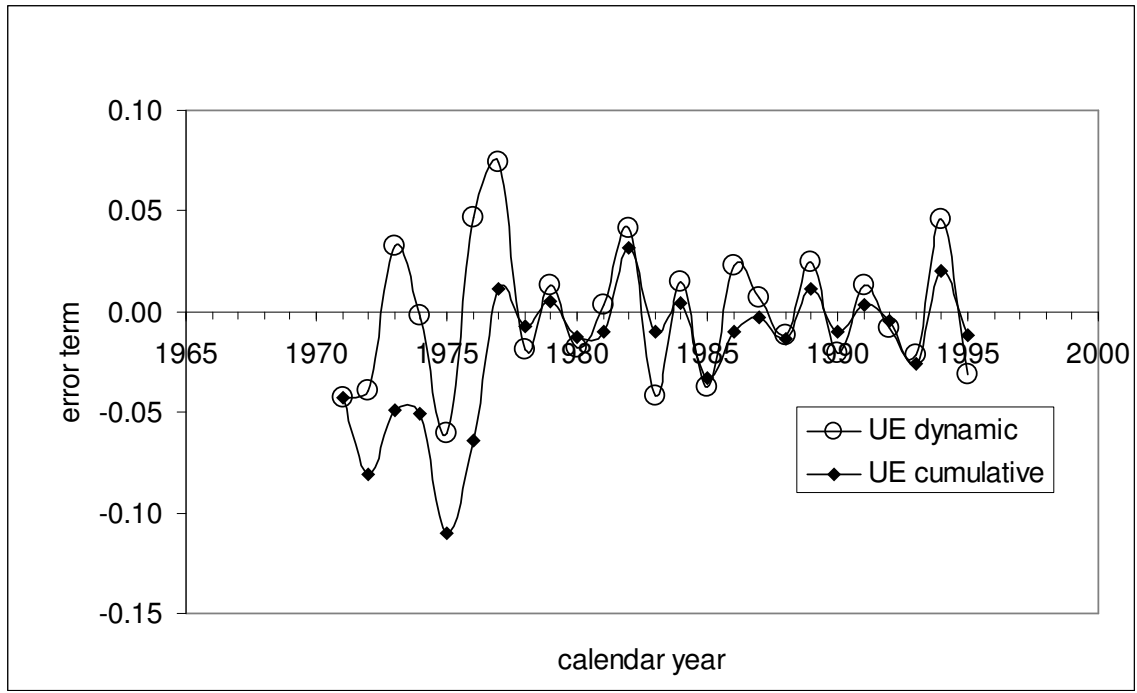

b)

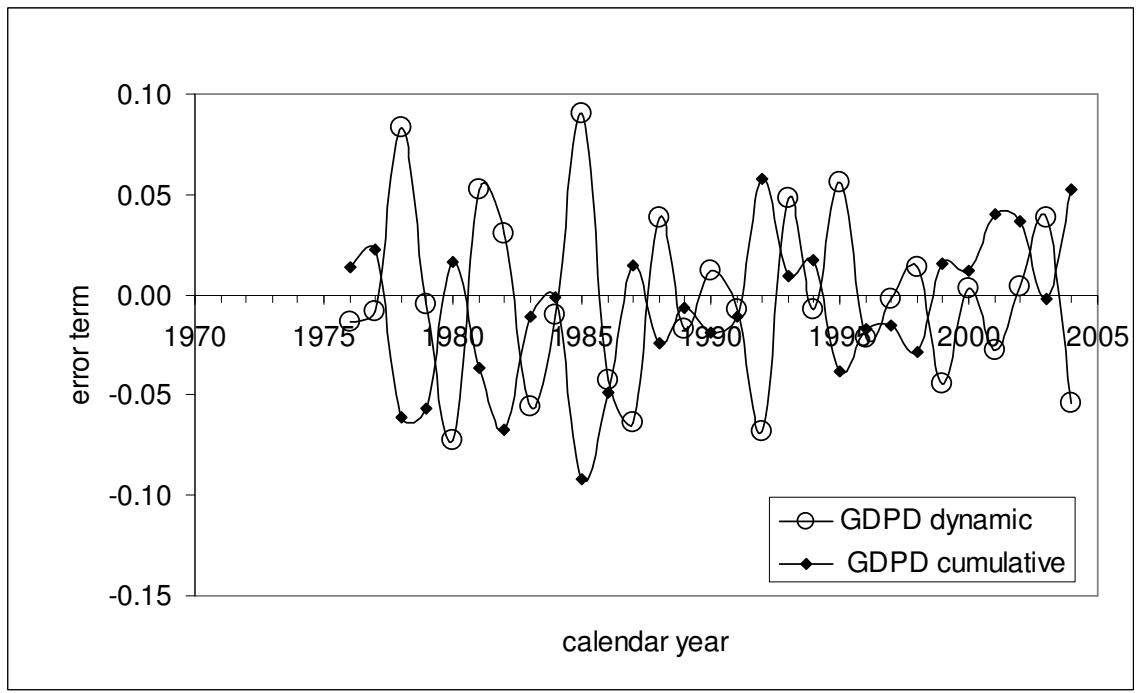

c)

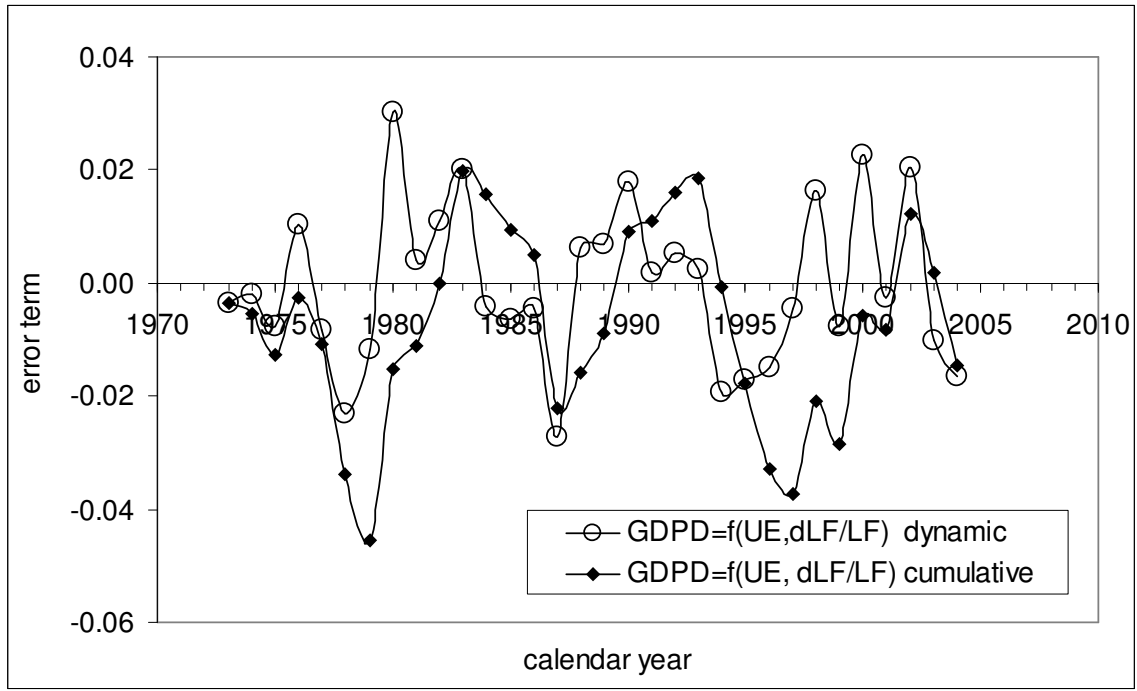

Figure 10. Comparison of residuals in the dynamic and cumulative relationships: a) *UE* vs. *dLF/LF*; b) *GDPD* vs. *dLF/LF*; c) *GDPD* vs. *UE* and *dLF/LF*.